\documentclass[12pt]{article}
\usepackage[latin1]{inputenc}
\usepackage{a4wide}
\usepackage{amsmath}
\usepackage{amssymb}
\usepackage{latexsym}
\usepackage{amsthm}
\usepackage{url}
\title{Feasible reactivity in a synchronous
  $\pi$-calculus\thanks{Work partially supported by ACI CRISS and ANR-06-SETI-010-02.}}
\author{
  Roberto M. Amadio\\
  Universit\'e Paris 7
  \thanks{Laboratoire {\em Preuves,Programmes et Syst\`emes},     UMR-CNRS 7126.} 
  \and Fr\'ed\'eric Dabrowski \\ INRIA Sophia-Antipolis
}

\newtheorem{definition}{Definition}
\newtheorem*{definition*}{Definition}

\newtheorem*{proposition*}{Proposition}
\newtheorem{lemma}{Lemma}
\newtheorem*{lemma*}{Lemma}
\newtheorem{theorem}{Theorem}
\newtheorem*{theorem*}{Theorem}

\newtheorem*{corollary*}{Corollary}
\newtheorem{remark}{Remark}
\newtheorem{example}{Example}

\newcommand{\ul}[1]{\underline{#1}}     

\newcommand{\isum}[3]{\Sigma_{i=1}^n}
\newcommand{\mset}[1]{\{\!|#1|\!\}}
\newcommand{\emit}[2]{\ol{#1}#2} 
\newcommand{\present}[4]{#1(#2).#3,#4}
\newcommand{\match}[4]{[#1=#2]\ #3,#4}
\newcommand{\matchv}[4]{[#1\unrhd #2]\ #3,#4}  
\newcommand{\new}[2]{\nu #1 \ #2}

\newcommand{\hide}{{\tt{0}}}

\newcommand{\w}[1]{{\it #1}}  
\newcommand{\ol}[1]{\overline{#1}}      
\newcommand{\s}[1]{{\sf #1}}    
\newcommand{\cl}[1]{{\cal #1}}          
\newcommand{\hatt}[1]{#1^{+}}
\newcommand{\vc}[1]{{\bf #1}}
\newcommand{\union}{\cup}               
\newcommand{\Union}{\bigcup}            
\newcommand{\minus}{\backslash}         
\newcommand{\set}[1]{\{#1\}}            
\newcommand{\cp}{\cl{C}}

\newcommand{\Call}[1]{{\it Call}(#1)}
\newcommand{\spi}{S\pi}
\newcommand{\Alt}{ \mid\!\!\mid  }
\newcommand{\arrow}{\rightarrow}        
\newcommand{\arrowr}{-\!\!\triangleright \ }
\newcommand{\trarrow}{\stackrel{*}{\rightarrow}} 
\newcommand{\eval}{\Downarrow}
\newcommand{\infer}[2]{\begin{array}{c} #1 \\ \hline #2 \end{array}}
\newcommand{\Defitem}[1]{\smallskip \noindent $#1\;$}
\newcommand{\Defitemt}[1]{\smallskip \noindent {\em #1\;}}

\newcommand{\susp}{\downarrow}
\newcommand{\real}{\makebox[5mm]{\,$\|\!-$}}

\newcommand{\act}[1]{\stackrel{#1}{\rightarrow}} 
\newcommand{\mapstoin}[1]{\stackrel{#1}{\mapsto}} 

\newcommand{\Nat}{\mathbb{N}}

\def\regs{\mathcal{R}}

\newcommand{\infr}[1]{\downarrow #1} 

\begin{document}
\maketitle
\begin{abstract}
Reactivity is an essential property of a synchronous
program. Informally, it guarantees that at each instant the program
fed with an input will `react' producing an output.  In the present
work, we consider a refined property that we call {\em feasible
reactivity}. Beyond reactivity, this property guarantees that at each
instant both the size of the program and its reaction time are bounded
by a polynomial in the size of the parameters at the beginning of the
computation and the size of the largest input.
We propose a method to annotate programs and we develop related static
analysis techniques that guarantee feasible reactivity for programs
expressed in the $\spi$-calculus.  The latter is a synchronous version of
the $\pi$-calculus based on the SL synchronous programming model.
\end{abstract}

\section{Introduction}
Mastering the {\em computational complexity} of programs is an
important aspect of computer security with applications ranging from
embedded systems to mobile code and smartcards.
One approach to this problem is to monitor at run time the resource
consumption and to rise an exception when some bound is reached. A
variant of this approach is to instrument the code so that bounds are
checked at appropriate time. An alternative approach is to analyse
{\em statically} the program to guarantee that during the execution
it will respect certain resource bounds.  In other words, the first
approach performs a {\em dynamic} verification while the second relies
on a {\em static} analysis. As usual, the main advantage of the first
approach is its flexibility while the advantage of the second approach
is the fact that it does not introduce an overhead at run time and,
perhaps more importantly, that it allows an early detection of `buggy'
programs.  In this work, we will focus on the static analyses which
offer the more challenging problems while keeping in mind that the two
approaches are complementary.  For instance, static analyses may be
helpful in reducing the frequency of dynamic verifications.

When addressing the issue of resource control, there is a variety of
properties of a program that one may check.  Termination is probably
the first one that comes to mind. However, in the context of {\em
interactive programs}, this property should be refined into {\em
reactivity}. In general, the set of reactive programs
can be defined (co-inductively) as the {\em largest} set $R$ of programs 
that terminate and such  that each interaction with the environment 
leads to a program which is again in the set $R$.

If a program manipulates data values of {\em variable size} such as
lists, trees, graphs, $\ldots$ then the analysis can go beyond
reactivity and, for instance, it can establish that the program
reacts while using a {\em feasible} amount of resources where feasible
can be understood, for instance, as computable in {\em polynomial time}.
In this case, the analysis produces a {\em function} that bounds the time (or
space) needed for the reaction depending on the size of certain
parameters. 

There is a large collection of static analysis techniques (see, {\em
e.g.}, \cite{Cobham65,BC92,Hofmann02,Jones97,Leivant94,BMM01} that
allow to establish feasible reactivity of {\em functional} programs.
A common feature of these methods is the combination of traditional
termination methods with what could be called a {\em data-size flow
analysis}.  By this we mean a method to describe how the size of the
values computed by a program depends on the size of the values taken in
input.

In \cite{AD04,AD05}, we have started a research programme that aims at
extending this approach to a synchronous, concurrent programming
language.  In the present work, we focus in particular on the
$\spi$-calculus \cite{Amadio06}.  This is a {\em synchronous} version
of the $\pi$-calculus \cite{MPW92} which is based on the SL
(synchronous language) model \cite{BD95}.  The latter can be regarded
as a relaxation of the {\sc Esterel} model \cite{BG92} where the
reaction to the {\em absence} of a signal within an instant can only
happen at the next instant.  Various full fledged concurrent and
synchronous programming languages have been developed on top of the SL
model (see, {\em e.g.}, \cite{SchemeFT,MandelPouzetPPDP05}) and the
$\spi$-calculus can be regarded as a more refined model capturing some
essential aspects of those languages.


Our contribution includes (i) a methodology to annotate programs and 
(ii) related static analysis methods that guarantee feasible 
reactivity for {\em finite
control} programs expressed in the $\spi$-calculus.  

Programs come
with two kinds of {\em annotations} that concern {\em thread identifiers}
and {\em signals}.
A characteristic of synchronous programs is that each thread
performs some set of actions in a {\em cyclic way}.
A cycle is different from an instant in that it can span
several instants (possibly an unbounded number of them).
We require that a subset of the thread identifiers mark
the end of a cycle and the beginning of a new one.
This annotation has no effect on the operational semantics
but it is used to produce certain static conditions.
The first condition is what we call the {\em read once condition}. 
Informally, this condition requires that each thread within each
cycle can only read a finite number of signals. The technical
consequence of this restriction is that the behaviour of a thread
within an instant can be described as a {\em function} 
of its parameters and the (finitely many) values read within the same cycle.

Thread identifiers carry two additional annotations.
A basic goal is to show that each instant terminates.
We are then naturally lead to {\em compare} thread identifiers and their
parameters according to some suitable well-founded order.
For this reason we assume that each thread identifier is
annotated with a {\em status} that describes how its parameters
should be compared (typically, according to a lexicographic
or multi-set order).
Another important goal towards feasible reactivity, is to show that the
parameters of a thread are in a sense {\em non-size increasing}.  It
turns out that it is not always appropriate to consider {\em all} parameters
and therefore we require that we explicitly associate with each thread
identifier the (possibly proper) subset of parameters that should be
considered in the analysis of its size.
To summarise, a thread identifier has three kinds of
annotations:
one saying whether it marks the end of a 
cycle, another, that we call status, describing 
how its parameters have to compared for termination
analysis, and a final one specifying the subset of the parameters 
that are relevant to the computation of its size.

On one hand, a program should be allowed to emit values on a signal that
depend on values read on other signals. 
On the other hand, we want to avoid situations where, for example,
a program repeatedly reads a value on a signal and emits a larger
value on the same signal.
To address this issue, we assume that signal names are partitioned 
into a finite number of {\em regions} 
which are ordered.  More precisely, we refine the type
system so that signal types come with a region $\rho$ as in the type
$\w{Sig}_{\rho}(t)$. In other terms, the  type of a signal name 
explicitly carries the information on the region to which the 
signal name belongs.
Again, this annotation does not affect the operational semantics but
it is used in the generation of static conditions that guarantee
feasible reactivity. Informally, the condition states that
the size of a value emitted on a signal at region $\rho$
is bound by a function of the size of the values read from signals
of smaller regions.

Next, we move on to an informal description of the static conditions. 
First of all, we have to find an abstract way to describe the 
data-size flow of a program. 
To this end, we import and adapt the concept of {\em
quasi-interpretation} that has been proposed in the context of the
analysis of the computational complexity of {\em first-order
functional programs} \cite{BMM01,Amadio04}.
As a second step, we describe a method to associate with a
program a finite set of inequalities on first-order terms and prove
that whenever these inequalities are satisfied by a (polynomially
bounded) quasi-interpretation the program is feasibly reactive.  The
inequalities can be classified in three categories according to their
purpose which is to ensure: (1) the termination of each instant, (2)
that the size of the parameters of a thread at the beginning of each
cycle is non-size increasing, (3) that the size of the values computed
by a thread within a cycle is bounded by a polynomial in the size of
the parameters of the thread and the size of the values read on the
signals within the cycle.  Obviously, these inequalities depend on the
signal and thread annotations we described above.

The rest of the paper is organised as follows.
In section \ref{section-spi} we introduce the syntax of
the $\spi$-calculus along with some programming examples and
an informal comparison with the $\pi$-calculus.
In section \ref{redsem}, we provide the formal reduction semantics
of the $\spi$-calculus and we introduce the notion of feasible
reactivity.
In section \ref{annotation-sec}, we define the different kinds of thread and 
signal annotations mentioned above, we show how to associate a set of 
inequalities with an annotated program, and we introduce the notion
of assignment which provides an interpretation of 
the inequalities in terms of numerical functions.
A quasi-interpretation is a polynomially bounded assignment which
satisfies the inequalities. Our main result states 
that a program that admits a quasi-interpretation is feasibly
reactive. We devote section \ref{proof} to an outline of
the proof techniques leaving the details in an appendix.

\section{The $\spi$-calculus}\label{section-spi}
We introduce the syntax of the $\spi$-calculus along with some
programming examples and an informal comparison with the
$\pi$-calculus.

\subsection{Programs}\label{programs}
Programs $P,Q,\ldots$ in the $\spi$-calculus
are defined as follows:
\[
\begin{array}{ll}
P &::= 0 \Alt A(\vc{e}) \Alt \emit{s}{e} \Alt
\present{s}{x}{P}{K} 
\Alt \match{s_1}{s_2}{P_1}{P_2}
\Alt \matchv{u}{p}{P_1}{P_2}
\Alt \new{s}{P}
\Alt P_1\mid P_2 \\
K &::=A(\vc{r})
\end{array}
\]
We use the notation $\vc{m}$ for a vector $m_1,\ldots,m_n$, $n\geq 0$.
The informal behaviour of programs follows.
$0$ is the terminated thread. $A(\vc{e})$ is a (tail) recursive call
of a thread identifier $A$ with a vector $\vc{e}$ of expressions as argument;
as usual the thread identifier $A$ is defined by a unique equation
$A(\vc{x})=P$ such that the free variables of $P$ occur in $\vc{x}$.
$\emit{s}{e}$ evaluates the expression $e$ and emits its value on the
signal $s$.  $\present{s}{x}{P}{K}$ is the {\em present} statement
which is the fundamental operator of the SL model. If the values
$v_1,\ldots,v_n$ have been emitted on the signal $s$ 
then $\present{s}{x}{P}{K}$ evolves
non-deterministically into $[v_i/x]P$ for some $v_i$ ($[\_/\_]$ is our
notation for substitution).  On the other
hand, if no value is emitted then the continuation $K$ is evaluated at
the end of the instant.  $\match{s_1}{s_2}{P_1}{P_2}$ is the usual
matching function of the $\pi$-calculus 
that runs $P_1$ if $s_1=s_2$ and $P_2$, otherwise.
Here both $s_1$ and $s_2$ are free.
$\matchv{u}{p}{P_1}{P_2}$, matches $u$ against the pattern $p$.
We assume $u$ is either a variable $x$ or a value $v$ and $p$ has
the shape $\s{c}(\vc{p})$, where $\s{c}$ is a constructor and 
$\vc{p}$ is a vector of patterns.  We also assume that if $u$ is a
variable $x$ then $x$ does not occur free in $P_{1}$.
At run time, $u$ is always a {\em value} and we 
run $\sigma P_1$ if $\sigma$ is the 
substitution matching $u$ against $p$ if it exists, and $P_2$ otherwise.
Note that as usual the variables occurring in the pattern $p$
(including signal names) are bound.
$\new{s}{P}$ creates a new signal name $s$ and runs $P$.
$(P_1\mid P_2)$ runs in parallel $P_1$ and $P_2$.  The continuation $K$
is simply a recursive call whose arguments are either expressions
or values associated with signals at the end of the instant in
a sense that we explain below. We will also write 
$\s{pause}.K$ for $\new{s}{\present{s}{x}{0}{K}}$ with $s$ not free in $K$. 
This is the program that waits till the end of the instant and then
evaluates $K$.

\subsection{Expressions}\label{expressions}
The definition of programs relies on the following syntactic categories:
\[
\begin{array}{lll}
\w{Sig} &::= s \Alt t \Alt \cdots  &\mbox{(signal names)} \\
\w{Var} &::= \w{Sig} \Alt x \Alt y \Alt z \Alt \cdots   &\mbox{(variables)} \\
\w{Cnst} &::= \s{*} \Alt \s{nil} \Alt \s{cons} \Alt \s{c} \Alt \s{d} \Alt\cdots &\mbox{(constructors)} \\
\w{Val} &::= \w{Sig} \Alt \w{Cnst}(\w{Val},\ldots,\w{Val})
&\mbox{(values $v,v',\ldots$)}\\
\w{Pat} &::= \w{Var} \Alt \w{Cnst}(\w{Pat},\ldots,\w{Pat})
&\mbox{(patterns $p,p',\ldots$)} \\
\w{Fun} &::=f \Alt g \Alt \cdots &\mbox{(first-order function
  symbols)} \\
\w{Exp} &::= \w{Var} \Alt \w{Cnst}(\w{Exp},\ldots,\w{Exp}) \Alt 
                          \w{Fun}(\w{Exp},\ldots,\w{Exp}) 
&\mbox{(expressions $e,e',\ldots$)} \\
\w{Rexp} &::= {!\w{Sig}} \Alt \w{Var} \Alt 
\w{Cnst}(\w{Rexp},\ldots,\w{Rexp}) \Alt \\ &\quad \w{Fun}(\w{Rexp},\ldots,\w{Rexp})
&\mbox{(exp. with deref. $r,r',\ldots$)} 
\end{array}
\]
As in the $\pi$-calculus, signal names stand both for
signal constants as generated by the $\nu$ operator and signal
variables as in the formal parameter of the present operator.
Variables $\w{Var}$ include signal names as well as variables of other
types.  Constructors $\w{Cnst}$ include $\s{*}$, $\s{nil}$, and $\s{cons}$.
Values $\w{Val}$ are terms built out of constructors and signal names.
The {\em size of a value} $|v|$ is defined as $|s|=|\s{c}|=0$ if $\s{c}$ is
a constant, and $|\s{c}(v_{1},\ldots,v_{n})|=1+\Sigma_{i=1,\ldots,n}
|v_{i}|$ if $n\geq 1$.
Patterns $\w{Pat}$ are terms built out of constructors and variables
(including signal names).  
We assume first-order function symbols $f,g,\ldots$ whose behaviour
will be defined axiomatically.
Expressions $\w{Exp}$ are terms built out of variables, constructors,
and function symbols.
Finally, $\w{Rexp}$ are expressions that may include 
the value associated with a signal $s$ at the
end of the instant (which is written $!s$, following the ML notation
for dereferenciation). Intuitively, this value
is a {\em list of values} representing the set of values emitted on 
the signal during the instant.
If $P, p$ are a program and a pattern then  we denote
with $\w{fn}(P), \w{fn}(p)$ the set of free signal names
occurring in them, respectively. We also use $\w{FV}(P), \w{FV}(p)$
to denote the set of free variables (including signal names).

\subsection{Typing}
Types include  the basic type $1$ inhabited by the constant $*$ and, 
assuming $t$ is a type,  the type $\w{Sig}(t)$ of signals carrying
values of type $t$, and the type $\w{list}(t)$ of lists of values of
type $t$ with constructors \s{nil} and \s{cons}.
In the examples, it will be convenient to abbreviate 
$\s{cons}(v_{1},\ldots,\s{cons}(v_{n},\s{nil})\ldots)$ with
$[v_{1};\ldots;v_{n}]$.
$1$ and $\w{list}(t)$ are examples of {\em inductive types}. More
inductive types (booleans, numbers, trees,$\ldots$) 
can be added along with more constructors.  
We assume that variables (including signals), constructor
symbols, and thread identifiers come with their (first-order) types. 
For instance, a constructor $\s{c}$ may have a type
$(t_1,t_2)\arrow t$ meaning that it waits two arguments of
type $t_1$ and $t_2$ respectively and returns a value of type $t$.
It is then straightforward to define when a program is well-typed 
and verify that this property is preserved by the following 
reduction semantics. 
We just notice that if a signal name $s$ has type $\w{Sig}(t)$ then its
dereferenced value $!s$ should have type $\w{list}(t)$. In the
following, we will tacitly assume that we are handling well typed
programs, expressions, substitutions,$\ldots$

\subsection{Comparison with the $\pi$-calculus}
The syntax of the $\spi$-calculus is similar to the one of
the $\pi$-calculus, however there are some important {\em semantic}
differences to keep in mind.

\Defitemt{Deadlock vs. End of instant.}
What happens when all threads are either terminated or
waiting for an event that cannot occur? In the $\pi$-calculus,
the computation stops. In the $\spi$-calculus (and more generally,
in the SL model), this  situation is detected and marks 
the end of the current instant.
Then suspended threads are  reinitialised, signals are 
reset, and the computation  moves to the following instant.

\Defitemt{Channels vs. Signals.}
In the $\pi$-calculus, a message is consumed by its recipient.
In the $\spi$-calculus, a value emitted along a 
signal persists within an  instant and it is reset at the end of it.
We note that in the semantics the only relevant information is 
whether a given value was emitted or not, {\em e.g.},
we do not distinguish the situation where the same value is emitted
once or twice within an instant.

\Defitemt{Data types.}
The (polyadic) $\pi$-calculus has {\em tuples} as basic data type, 
while the $\spi$-calculus has {\em lists}. 
The reason for including lists rather than tuples in the
{\em basic} calculus is that at the end of the instant we transform a
set of values into a suitable data structure (in our case a list) that
represents the set and that can be processed as a whole 
in the following instant. Note in particular, that the list associated
with a signal is \s{nil} if and only if no value was emitted on the
signal during the instant. This allows to detect the {\em absence} of 
a signal at the end of the instant.

We consider a simple example that illustrates our discussion.
Assume $v_1\neq v_2$ are two distinct values and consider the
following program in $\spi$:
\[
\begin{array}{l}
P=\nu \ s_1,s_2 \ 
(\quad \emit{s_{1}}{v_{1}} \quad \mid \quad 
 \emit{s_{1}}{v_{2}} \quad \mid  \quad 
 s_1(x). \ (s_1(y). \  (s_2(z). \ A(x,y) \  \ul{,B(!s_1)}) \quad
 \ul{,0}) \quad \ul{,0} \quad )
\end{array}
\]
If we forget about the underlined parts and we regard $s_1,s_2$ as
{\em channel names} then $P$ could also be viewed as a $\pi$-calculus
process. In this case, $P$ would reduce to 
\[
P_1 =  \new{s_1,s_2}{(s_2(z).A(\sigma(x),\sigma(y))}
\]
where $\sigma$ is a substitution such that
$\sigma(x),\sigma(y)\in \set{v_1,v_2}$ and $\sigma(x)\neq \sigma(y)$.
In $\spi$, {\em signals persist within the instant} and 
$P$ reduces to 
\[
P_2 = \new{s_1,s_2}{(\emit{s_{1}}{v_{1}} \mid 
 \emit{s_{1}}{v_{2}} \mid (s_2(z).A(\sigma(x),\sigma(y)),B(!s_1)))}
\]
where  $\sigma(x),\sigma(y)\in \set{v_1,v_2}$.
What happens next? In the $\pi$-calculus, $P_1$ is 
{\em deadlocked} and no further computation is possible.
In the $\spi$-calculus, 
the fact that no further computation is possible in $P_2$ is
detected and marks the {\em end of the current instant}. Then
an additional computation represented by the relation $\mapsto$ 
moves $P_2$ to the following instant:
\[
P_2 \mapsto P'_2 =  \new{s_1,s_2}{B(v)}
\]
where $v \in  \set{[v_1;v_2],[v_2;v_1]}$.
Thus at the end of the instant, a dereferenced signal such as $!s_{1}$
becomes a list of (distinct) values emitted on $s_1$ during the
instant and then all signals are reset.

\subsection{Programming examples}\label{programming-example}
We introduce a few programming examples on which we will
rely in the following to illustrate our static analysis techniques.

\begin{example}\label{cellular-automaton-ex}
The synchronous model is particularly adapted to the 
{\em simulation} of various kinds of systems
(we refer to \cite{mimosarp} for a number of examples).
Here,  we describe the behaviour of a cell of a generic
cellular automaton. Each cell relies on three 
parameters: its own activation signal $s$, its state $q$, and
the list $\ell$ of activation signals of its neighbours.
The cell performs the following operations in a cyclic
fashion: 
(i) it emits its current state on the activation signals
of its neighbours,
(ii) it suspends for the current instant, and
(iii)  it collects the values emitted by its neighbours 
and computes its new state.
This behaviour can be programmed as follows:
\[
\begin{array}{lcll}
  \w{Cell}(s,q,\ell)&=& \w{Send}(s,q,\ell,\ell) \\
  \w{Send}(s,q,\ell,\ell')   &=&
  \matchv{\ell'}{\s{cons}(s',\ell'')} 
  {&(\emit{s'}q \mid \w{Send}(s,q,\ell,\ell''))}
  {\\&&&\s{pause}.\w{Cell}(s,next(q,!s),\ell)}
\end{array}
\]
where $\w{next}$ is a function that computes the following state
of the cell according to its current state and the
state of its neighbours.
We assume some finite enumerated type `\w{state}' that contains a constant
for each state. 
The type of the signals $s,s'$ is $\w{Sig}(\w{state})$,
 the type of the lists $\ell,\ell'$ is $\w{list}(\w{Sig}(\w{state}))$, 
and the type of the function $\w{next}$ is $\w{state},\w{list}(\w{state})\arrow \w{state}$.
\end{example}

\begin{example}\label{server-ex}
This example describes a `server' handling a list of requests
emitted in the previous instant on the signal $s$. 
For each request of the shape $\s{req}(s',x)$, 
it provides an answer which is a function of $x$ along the signal $s'$.
\[
    \begin{array}{lcl}
      \w{Server}(s)&=&{\tt{pause}}.\w{Handle}(s,!s)\\
      \w{Handle}(s,\ell)&=&
      \matchv{l}{\s{cons}(\s{req}(s',x),\ell')}
      {(\emit{s'}{f(x)} \mid \w{Handle}(s,\ell'))}
      {\w{Server}(s)}
    \end{array}
\]
Assume the function $f$ has type $t\arrow t'$ and assume 
an inductive type $\w{treq}$ with a constructor
$\s{req}: \w{Sig}(t'),t \arrow \w{treq}$.
Then the parameters $s$ have type $\w{Sig}(\w{treq})$ and
the lists $\ell,\ell'$ have type $\w{list}(\w{treq})$.
\end{example}

\begin{example}\label{ABC-ex}
This example describes two threads: the thread $A(s)$ re-emits 
on $s$ the values that were emitted
on $s$ in the previous instant while the thread $C(s)$ emits a (fresh) value on
$s$.
\[
\begin{array}{ll}

A(s)   &= \s{pause}.B(s,!s) \\
B(s,\ell) &= 
\matchv{\ell}{\s{cons}(n,\ell')}{(\emit{s}{n} \mid B(s,\ell'))}{A(s)}\\
C(s)      &=\new{n}{\emit{s}{n} \mid \s{pause}.C(s)}

\end{array}
\]
Assuming $n$ has type $\w{Sig}(1)$, $s$ has type $\w{Sig}(\w{Sig}(1))$, 
and the list $\ell$ has type $\w{list}(\w{Sig}(1))$.
\end{example}

\section{Reduction semantics and feasible reactivity}\label{redsem}
We provide the formal reduction semantics
of the $\spi$-calculus and we introduce the notion of feasible
reactivity.

\subsection{Expression evaluation}
We assume an evaluation relation $\eval$ such that for every
function symbol $f$ and values $v_1,\ldots,v_n$ of suitable type
there is a unique value $v$ such that $f(v_1,\ldots,v_n)\eval v$,
$\w{fn}(v)\subseteq \Union_{i=1,\ldots,n}\w{fn}(v_{i})$,
and moreover we suppose that the value $v$ can be computed in time
polynomial in the size of the values $v_1,\ldots,v_n$.
As already mentioned, the techniques for defining 
first-order functional programs
that enjoy these properties are well-studied. 
The evaluation relation $\eval$ is extended to expressions as usual:
$$
\infer{}{s\eval s}
\qquad
\infer{e_i\eval v_i \quad i=1,\ldots,n}
{\s{c}(e_1,\ldots,e_n)\eval \s{c}(v_1,\ldots,v_n)} 
\qquad
\infer{e_i\eval v_i \quad i=1,\ldots,n \quad f(v_1,\ldots,v_n)\eval v}
{f(e_1,\ldots,e_n)\eval v}
$$
We will abbreviate $e_1\eval v_1,\ldots,e_n\eval v_n$ with
$\vc{e}\eval \vc{v}$.

\subsection{Reduction semantics}\label{red-sem-sec}
The (internal) behaviour of a program is specified by (i) a {\em reduction
system} $\arrow$ describing the possible reductions of the program
{\em during an instant} and (ii) an {\em evaluation relation} $\mapsto$ determining
how a program evolves at the {\em end of each instant}. 
These definitions rely on a structural equivalence relation $\equiv$
that we introduce first.

\subsubsection{Structural equivalence}
The {\em structural equivalence} $\equiv$ is the least equivalence relation 
on programs that identifies programs up to $\alpha$-renaming and that
satisfies the following standard equations:
$$
\begin{array}{c}
P\mid 0 \equiv P,
\qquad

P_1 \mid P_2 \equiv P_2 \mid P_1, 
\qquad
(P_1 \mid P_2) \mid P_3 \equiv P_1 \mid (P_2 \mid P_3),  \\

\new{s}{P} \equiv P \mbox{ if }s\notin\w{fn}(P),
\qquad

\new{s}{P_1 \mid P_2} \equiv \new{s}{(P_1\mid P_2)} 
\mbox{ if }s\notin\w{fn}(P_2) ~.

\end{array}
$$

\subsubsection{Reduction relation}
We introduce the following {\em reduction rules}:
$$
\begin{array}{cc}
\infer{e\eval v}
{\emit{s}{e} \mid \present{s}{x}{P}{K} \arrowr  \emit{s}{e} \mid  [v/x]P} 
\quad

&\infer{A(\vc{x})=P \quad \vc{e}\eval \vc{v}}
{A(\vc{e}) \arrowr [\vc{v}/\vc{x}]P} \\ \\ 

\infer{~}
{[s=s]P_1,P_2 \arrowr P_1}

&\infer{s\neq s'}
{[s=s']P_1,P_2 \arrowr P_2} \\ \\ 

\infer{\w{match}(v,p)=\sigma}
{\matchv{v}{p}{P_1}{P_2} \arrowr \sigma P_1}

&\infer{\w{match}(v,p) \mbox{ undefined}}
{\matchv{v}{p}{P_1}{P_2} \arrowr P_2} 

\end{array}
$$
A {\em static context} $C$ is defined by
$C::= [~] \Alt \new{s}{C} \Alt (C\mid P)$.
The {\em reduction relation} $\arrow$ is then defined by the rule:
\[
\infer{P\equiv C[P']\quad P'\arrowr Q'\quad C[Q']\equiv Q}{P\arrow Q}
\]

\subsubsection{Suspension and evaluation at the end of the instant}\label{eoi-def}
We write $P\susp$ if $\neg\exists Q\ (P\arrow Q)$ and say that the
program $P$ is {\em suspended}.  When $P$ is suspended the instant
ends and an additional computation is carried on to move to the next
instant. This goes in three steps that 
amount to:
(1) collect in lists the set of values emitted on every signal, 
(2) extrude the signal names contained in values visible at the end of
the instant, and
(3) initialise the continuations $K$ of the present statements.

To this end, we introduce first some notation.  A suspended program $P$ is
structurally equivalent to:
\begin{equation}\label{susp}
\nu \vc{s} (S \mid \w{In})
\end{equation}
where the signal names $\vc{s}$ are all distinct, 
$S\equiv \emit{s_{1}}{e_{1}} \mid \cdots \mid \emit{s_{n}}{e_{n}}$, 
$\w{In} \equiv t_1(x_1).P_1,A_1(\vc{r_{1}}) \mid \cdots
\mid t_m(x_m).P_m,A_m(\vc{r_{m}})$, and $n,m\geq 0$
(by convention an empty parallel composition equals the program $0$).
We write $\emit{s}{e}\in S$ to mean that $\emit{s}{e}$ occurs
in the parallel composition $S$. 
We can now formalise the steps (1--3).

\begin{enumerate}

\item Let $V$ be a function from signal names to lists of values.  We
say that $V$ {\em represents} $S$ and write $V\real S$ if for all signal
names $s$, if $\set{v_1,\ldots,v_n}=\set{v \mid \emit{s}{e}\in S,\ e\eval v}$
then
$V(s)=[v_{\pi(1)};\cdots;v_{\pi(n)}]$
for some permutation $\pi$.

\item 
We define $\w{Free}(\nu\vc{s} \ S)$ as the least set of signal names
such that $\w{Free}(\nu\vc{s} \ S)\supseteq \w{fn}(\nu\vc{s} \ S)$ and
if $s\in \w{Free}(\nu\vc{s} \ S)$, $\emit{s}{e}\in S$, $e\eval v$, and 
$s'\in \w{fn}(v)$ then $s'\in \w{Free}(\nu\vc{s} \ S)$.
For instance, $\w{Free}(\nu s_1,s_2 \ \emit{s}{s_{1}} \mid
\emit{s_{1}}{s_{2}}) = \set{s,s_1,s_2}$.

\item If $r$ is an expression with dereferenciation then
$V(r)$ is the expression resulting from the replacement of
all dereferenced signals $!s$ with $V(s)$.
If $A(\vc{r})$ is a continuation $K$ of a present statement,
where $\vc{r}$ are closed expressions,
then $\w{Eval}(A(\vc{r}),V)= A(\vc{v})$
if $V(\vc{r}) \eval \vc{v}$.
Finally, if $\w{In}$ is defined as in (\ref{susp}) then
$\w{Eval}(\w{In},V) = \w{Eval}(A_1(\vc{r}_{1}),V) \mid \cdots \mid 
\w{Eval}(A_m(\vc{r}_{m}),V)$.

\end{enumerate}

With these conventions, we can now state the evaluation rule at the
end of the instant:
\[
\infer{P\susp \quad P \equiv \nu \vc{s} \ (S\mid \w{In}) \quad 
V\real S \quad \set{\vc{s'}} = \set{\vc{s}}\minus \w{Free}(\nu \vc{s}
\ S) \quad P'\equiv \nu \vc{s'} \w{Eval}(\w{In},V)}
{P \mapsto P'}
\]
In this rule, (i) we decompose the suspended program 
in emissions and inputs, (ii) we compute a
representation of the emission, (iii) we compute the signal names
extruded, and finally (iv) we remove the emitted names and 
initialise the continuations of the present statements.

\subsection{Feasible reactivity}
At the beginning of each instant, a program receives
an input that we may represent as a (fresh) thread identifier
$\w{Env}$ defined by the equation
$\w{Env}()=\emit{s_{1}}{v_{1}} \mid \cdots \mid \emit{s_{n}}{v_{n}}$.
Then we write
\[
P\mapstoin{\w{Env}} P' \mbox{ if }P\mapsto P'' \mbox{ and } P'\equiv (P''\mid \w{Env})
\]
By the properties of the model, we may assume without loss of generality 
that in the input all values emitted on a signal $s$ 
are {\em distinct}.  

\begin{definition}[computation]
A computation of a program $P$ is an infinite and countable 
sequence of programs $P_1,P_2,\ldots$  such that
\[
P \equiv P_{1} \trarrow  P_{i_{1}} \mapstoin{\w{Env}_{1}} P_{i_{1}+1} \trarrow
P_{i_{2}}  \mapstoin{\w{Env}_{2}} P_{i_{2}+1} \cdots
\]
\end{definition}

In general, the reduction of $P_1,P_{i_{1}+1},P_{i_{2}+1},\ldots$ may fail to reach the
end of the instant. We call {\em reactive} the programs that
are guaranteed to suspend.

\begin{definition}[reactivity]
A program $P$ is reactive if in all computations that
start with $P$, the evaluation at the end of the instant occurs
infinitely often.
\end{definition}

\begin{example}
With reference to the example \ref{ABC-ex},
a possible computation of the program $A(s)\mid C(s)$ is as
follows:
\[
\begin{array}{lll}

A(s) \mid C(s) 
&\trarrow \s{pause}.B(s,!s) \mid \new{n_{0}}{\emit{s}{n_{0}}} \mid \s{pause}.C(s)  
&\mapstoin{\w{Env}_1} B(s,[n_{0}]) \mid C(s) \\
&\trarrow \emit{s}{n_{0}}\mid \s{pause}.B(s,!s) \mid \new{n_{1}}{\emit{s}{n_{1}}} \mid \s{pause}.C(s) 
&\mapstoin{\w{Env}_2} B(s,[n_{0}; n_{1}]) \mid C(s) \cdots 

\end{array}
\]
In this case, we assume that
the input at the beginning of each instant is empty, $\w{Env}_{i}()=0$
for $i=1,2,\ldots$.
Note that the order of the signal names in the list $\ell$, which is a
parameter of the identifier $B$, is chosen non-deterministically at the
beginning of each instant.
\end{example}

We assume that initially a program has the shape 
\begin{equation}\label{shape-init}
\nu \vc{s}( A_1(\vc{v_{1}}) \mid \cdots \mid A_n(\vc{v_{n}}) )
\end{equation}
Then, by the definition of the present instruction and the input, a program
will have this shape at the beginning of each instant, up to structural
equivalence.
The definition of feasible reactivity is relative to the
size of the initial program and the size of the (largest) input.
By convention, the size of a program with the shape
(\ref{shape-init}) is $n$ plus the sum of the sizes
of the values $\vc{v_{1}},\ldots, \vc{v_{n}}$.
The size of an input $\w{Env}$ defined by an equation
$\w{Env}=\emit{s_{1}}{v_{1}}\mid
\cdots \mid \emit{s_{n}}{v_{n}}$ is the size of the list
$[v_{1};\ldots;v_{n}]$.


\begin{definition}[feasible reactivity]\label{feas-react-def}
A program $P$ of the shape (\ref{shape-init}) 
is  {\emph{feasibly reactive}} if there exists a 
polynomial $Q$ such that for every computation
\[
P \equiv P_{i_{0}+1} \trarrow  P_{i_{1}} \mapstoin{\w{Env}_{1}} P_{i_{1}+1} \trarrow
P_{i_{2}}  \mapstoin{\w{Env}_{2}} P_{i_{2}+1} \cdots
\]
if $d$ bounds the size of $P$ and the sizes of $\w{Env}_{1},\ldots,\w{Env}_{k}$ 
for  $k\geq 1$ then (i) $P_{i_{k}+1}$ (the program at the beginning of the
instant $k$) has size bounded by 
$Q(d)$ and (ii) it is guaranteed to suspend  in time less than $Q(d)$,
\end{definition}

For instance, the program in example \ref{ABC-ex} fails to be feasibly
reactive because the size of the parameter $\ell$ of the identifier
$B$ grows by one every instant.

\section{Annotations and Constraints Generation}\label{annotation-sec}
Programs come with a finite system of recursive equations. Our static
analysis actually concerns this system and it is independent of the
particular program that is used to initialise the computation. The
reader should keep in mind that the analysis of a program is actually
the analysis of the associated system.
We restrict our attention to {\em finite control programs}. To this
end, we inspect the system of equations and we check that in each
equation $A(\vc{x})=P$, $P$ cannot spawn two recursive calls that run
in parallel.
Also, the static analysis makes abstraction of the actual signal names
while keeping track of the region they belong to. It will be
convenient to suppose that the program does not contain trivial
matchings such as a value matching a pattern
($\matchv{v}{p}{P_{1}}{P_{2}}$) and the comparison of two identical
names ($\match{s}{s}{P_{1}}{P_{2}}$).  Such matchings can be removed
by a trivial symbolic execution.

\subsection{Reset annotations and read once condition}\label{read-once}
We denote with $\w{Reset}$ a subset of the thread identifiers 
containing those thread identifiers that correspond to the beginning of
a new `cycle'.
To be in $\w{Reset}$ a thread identifier $A$ has to satisfy one of
the following conditions: either it is defined by an equation
of the shape $A(\ldots)=\s{pause}.K$ or all its occurrences
in the program are in the else branch of a \s{present} statement.
By these syntactic conditions, we guarantee that the end of a cycle for
a given thread always entails the end of its computation 
for the current instant. 
For instance, in the example \ref{server-ex}, it is natural to assume that
$\w{Server}\in \w{Reset}$ and $\w{Handle}\notin \w{Reset}$.

As we have seen, a program may {\em read} a signal during an instant
with the present statement or at the end of the instant through
dereferencing. The {\em read once condition} is  the hypothesis that 
for every thread, in every cycle, 
there is a bound on the number of times the reading
of a signal can be performed. Specifically, we require and
statically check on the \emph{call graph} of the program
(see below) that the computation performed starting from
any thread identifier can execute any given read
instruction at most once within a cycle.

\begin{enumerate}  
\item
  We assign to every present statement and to every
  dereferencing  in a program a distinct fresh
  label (a variable), $y$, and we collect all these labels in an ordered sequence,
  $y_1,\ldots, y_m$. In the following, we will use the
  notation $\present{s^{y}}{x}{P}{K}$ and $!^y s$ to make 
  the labels explicit. 
  If $\vc{r}$ is a vector of expressions with dereferenciation, we denote 
  with $\w{Lab}(\vc{r})$ the finite set of labels that occur in $\vc{r}$.

\item   With every thread identifier $A$ defined by an equation
  $A(\vc{x})=P$, we associate a node of the graph. We also introduce
a fresh thread identifier $O$ and a related node that plays the role
of a {\em sink} in the call graph.

\item We define a function $\w{Call}$ that takes in input
a program and a finite set of labels and produces in output
a finite set of pairs composed of a thread identifier and 
a set of labels. 
The function \w{Call} is defined as follows:
  $$
  \begin{array}{c}
    \begin{array}{lcl}    
      \Call{0,L}&=&\set{(O,L)}\\

      \Call{\emit{s}{e},L}&=&\set{(O,L)}\\

      \Call{\present{s^y}{x}{P}{A(\vc{r})},L} &=& 
       \left\{
     \begin{array}{ll}
      \Call{P,L\union\set{y}} 
      \union \{(A,L\union \w{Lab}(\vc{r})) \}
      &{\text{if }} A\notin \w{Reset} \\
      \Call{P,L\union\set{y}} \union  \{(O,L\union \w{Lab}(\vc{r}))\}
      &  {\text{ otherwise}}
      \end{array} \right.  \\
      \Call{A(\vc{e}),L} &=&
\left\{
     \begin{array}{ll}
       \set{(A,L)}       &{\text{if }} A\notin \w{Reset} \\
      \set{(O,L)}         &  {\text{ otherwise}}
      \end{array} \right.  \\
      \Call{\match{s_1}{s_2}{P_{1}}{P_{2}},L} &=& \Call{P_{1},L}
      \union \Call{P_{2},L}\\
      \Call{\matchv{x}{p}{P_{1}}{P_{2}},L} &=& \Call{P_{1},L}
      \union \Call{P_{2},L}\\
      \Call{P_1\mid P_2,L}&=&\Call{P_{1},L}\union \Call{P_{2},L}\\
      \Call{\new{s}{P},L}&=&\Call{P,L}
    \end{array}
  \end{array}
  $$

\item Suppose the identifier $A$ is defined by an equation
$A(\vc{x})=P$ and that $C=\Call{P,\emptyset}$. 
We introduce an edge from $A$ to an identifier $B$ (possibly $O$)
if $(B,L)\in C$. In this case, we label the edge with
the set $\Union \set{L \mid (B,L)\in C}$.

\item We denote with $R(A)$ the union of the sets of labels 
of the edges accessible from $A$ and  with $\vc{y}_A$
the ordered sequence of labels in $R(A)$.  
\end{enumerate}

The definition of \w{Call} is such that for every sequence of calls in
the execution of a thread within the cycle we can find a
corresponding path in the call graph.

\begin{definition}[read once condition]
  A program satisfies the read once condition if in the call graph
  there are no loops that go through an edge whose label is a non-empty set.
\end{definition}

Note that while the number of reads is bounded by a constant, the
amount of information that can be read is not. Thus, for instance,
a `server' thread can just read one signal in which is stored the list
of requests produced so far and then it can go on scanning the list
and replying to all the requests within the same instant.
In the following, we will focus on programs that satisfy the read
once condition.
For such programs, we introduce for each thread identifier
$A$ with parameters $\vc{x}$, a fresh thread identifier $\hatt{A}$
whose parameters are those of $A$ plus the parameters
$\vc{y}_A$ that can be read within a cycle.
The idea is that the behaviour generated by the thread identifier
$A$ within a cycle can be described as a {\em function} of its
parameters $\vc{x}$ which are determined at the beginning of
the cycle and the values $\vc{y}_A$ of the signals read within
the cycle. We will also refer to $\vc{x}$ as {\em proper parameters}
and to $\vc{y}_{A}$ as {\em auxiliary parameters} of the identifier $\hatt{A}$.

\begin{example}\label{call-graph-ex}
Consider example \ref{cellular-automaton-ex} and suppose
that $\w{Cell}\in \w{Reset}$ marks the end of a cycle and
that the label associated with the dereferenciation is $y$. 
The graph resulting from the
analysis has three nodes $\set{\w{Cell},\w{Send},O}$ and 
the following labelled edges: $(\w{Cell},\emptyset,\w{Send})$,
$(\w{Send},\emptyset, \w{Send})$ and $(\w{Send},\set{y},O)$.
The program satisfies the read once condition since
the only possible loop, namely the one form \w{Send} to \w{Send},
is composed of edges (just one in this case) whose label is the empty set.
Both $\hatt{\w{Send}}$ and $\hatt{\w{Cell}}$ have an auxiliary parameter
$y$.

Next consider example \ref{server-ex} and suppose that $\w{Server}\in \w{Reset}$
and the label associated with the dereferenciation is $y$.
The call graph has three nodes
$\w{Server},\w{Handle},O$ and the following labelled edges:
$(\w{Server},\set{y},\w{Handle})$, $(\w{Handle},\emptyset,\w{Handle})$, 
and $(\w{Handle},\emptyset,O)$. Again the read once condition is
satisfied. $\hatt{\w{Server}}$ has an additional parameter $y$ while
$\hatt{\w{Handle}}$ has no additional parameter.

Finally, consider example \ref{ABC-ex} and suppose that 
$A,C\in \w{Reset}$ and the label associated with the dereferenciation 
is $y$. The graph has four nodes: $C,A,B,O$ and the following
labelled edges:
$(A,\set{y},B)$, $(B,\emptyset,B)$, $(B,\emptyset,O)$, and $(C,\emptyset,C)$.
In this case too, the read once condition is satisfied. $\hatt{A}$ has
an additional parameter $y$ while $\hatt{B}$ and $\hatt{C}$ have no additional
parameter.
\end{example}

\subsection{Status annotations}
We associate a \emph{status}, either \emph{lexicographic} ($lex$)
or \emph{multi-set} ($mset$), with every thread identifier.
We assume that thread identifiers which are equivalent 
with respect to a pre-order $\geq_F$ that we define below 
have the same arity and the same status.
We note that this implies that $\hatt{A},\hatt{B}$ have
the same arity too.

To define the pre-order $\geq_F$, 
we introduce first a {\em call graph within an instant}
by modifying the definition given in section 
\ref{read-once}  so that
$\w{Call}(A(\vc{e}),L)=\set{(A,L)}$ and
$\w{Call}(s(x).P,K,$ $L) = \w{Call}(P,L)$.
Thus there is an edge from the identifier $A$ to the identifier
$B$ if in the definition of $A$, say $A(\vc{x})=P$, 
it is possible to call $B$ within the same instant $A$ is called.
Second, we  build the least pre-order
(reflexive and transitive) $\geq_F$ over thread identifiers
such that $A \geq_F B$ if there is an edge from $A$ 
to $B$ in the call graph within an instant.
We write $A=_F B$ if $A\geq_F B$ and $B\geq_F A$, and
$A>_F B$ if $A\geq_F B$ and $A\not=_{F} B$.
The rank of the thread identifier $A$, noted $rank(A)$,
is the length of the longest chain $A >_F B >_F \ldots$

\subsection{Parameter annotations}
One of our goals is to control the size of the proper parameters of a thread.
However, it is sometimes appropriate
to {\em neglect} some parameters. For instance, consider the example
\ref{server-ex}. 
One of the parameters of the thread identifier $\w{Handle}$ is a list
$\ell$ that is read on a signal $s$ whose size is unrelated to the size
of the parameter $s$ of the thread identifier $\w{Server}$.
We observe that the parameter $\ell$ is needed by $\w{Handle}$ to
perform some computation and that this parameter is then neglected
at the end of the cycle. We then introduce a mechanism to {\em mask}
parameters such as $\ell$.
Let $\hide$ be a fresh constant that stands for a parameter of size
$0$. If $h$ is a function
of arity $n$ and $I\subseteq \{1,\ldots,n\}$ is a subset of its 
parameters then $h(e_1,\ldots,e_n)_I$ is defined as $h(e_1',\ldots,e_n')$
where $e_i'=e_i$ if $i \in I$ and $e_i'=\hide$ otherwise.
Intuitively, in $h(e_1,\ldots,e_n)_I$ `we set to $0$' all
arguments that are not in $I$.
For each thread identifier $A$ defining a behaviour of arity $n$,
we assume a set $I_A\subseteq \{1,\ldots,n\}$ with the
condition that $I_A=\{1,\ldots,n\}$ if $A$ marks the end of a cycle in the program
(thus in the latter case, no parameter can be set to $0$).
Note that the mask acts only on the proper parameters of the identifier $A$
and {\em not} on the auxiliary parameters $\vc{y}_{A}$ corresponding
to the values read within a cycle.

\subsection{Signal annotations}
One purpose of the signal annotations is to reject programs such as
the one in example \ref{ABC-ex}. Let us consider in particular the
thread $A$.  At each instant, this thread re-emits on a signal $s$ the
values emitted on the {\em same} signal $s$ at the previous instant.
We want to reject this kind of behaviour while allowing --under
suitable conditions-- a slightly different behaviour where a thread
emits on a signal $s$ a series of values (possibly the same) that
depend on the values emitted on a {\em different} signal $s'$ at the
previous instant. For instance, we want to be able to program a
`server' (cf. example \ref{server-ex}) 
that receives a series of requests at the end of the instant
and produces a series of related answers in the following instant.
The idea is to partition the signal names into a finite collection of
{\em regions}. Then regions are ordered and the behaviour of the
server described above is allowed if the signal $s$ belongs to a
region that is strictly below the region to which $s'$ belongs.
For instance, in the example \ref{server-ex}, we emit on signal $s'$ a value
which depends on a value read on a signal $s$. If we admit
that this value has arbitrary size then we should require that
the signal $s$ is associated with a region smaller than the
region associated with $s'$.

Formally, we assume a set of regions
$\regs=\set{\rho_1,\rho_2,\ldots}$ with a strict order $>_\regs$
and we denote with $\w{rank}(\rho)$ the length of the longest
sequence $\rho >_\regs \rho_1 >_\regs \cdots >_\regs \rho_n$.
We assume that every signal type comes with a region annotation
$\w{Sig}_\rho(t)$ so that the type of a signal name also provides the
region to which the signal name belongs.
In section \ref{inequalities-sec}, we will
rely on these annotations to derive inequalities that guarantee that
the size of the values emitted on a signal of region $\rho$ can be
bound as a function of the size of the values received on signals
belonging to regions of smaller rank.

\subsection{Inequalities}
We rely on the annotations to produce a set of inequalities.
We use the notation  $\ol{\vc{r}}$ for $\vc{r}$ where 
each $!^y s$ is replaced with $y$.
Given a system of equations, for each thread identifier 
$A$ defined by an equation $A(\vc{x})=P$, we 
compute $\cl{C}_i(P,\hatt{A}(\vc{x},\vc{y}_A))$, with index $i=0,1,2$
according to the rules described in table \ref{inequalities}.
The definition of the functions $\cl{C}_i$ amounts to 
perform a `symbolic execution' of the body $P$ of the
equation while keeping track of the shape of the parameters
$\vc{x}$ and the values read $\vc{y}_A$. 
More precisely, the functions $\cl{C}_i$ explore the finitely many
control points of a computation starting with a recursive call
to the thread identifier $A$.
At some critical points, namely (i) when a value is emitted, 
(ii) when a value is received, and
(iii) when a recursive call is executed, the functions $\cl{C}_i$
produce certain inequalities whose purpose is discussed next.

\begin{table}
\[
\begin{array}{ll}
  \cl{C}_i(P,\hatt{A}(\vc{p}))
  &= \s{case} \ P \ \s{of} \\[1ex]
  0&:\emptyset \\[1ex]

  \matchv{x}{p}{P_1}{P_2}  &:\cl{C}_i(P_1,\hatt{A}([p/x]\vc{p})) \union 
  \cl{C}_i(P_2,\hatt{A}(\vc{p})) \\[1ex]

  [s_1=s_2]P_1,P_2   &:\cl{C}_i(P_1,\hatt{A}(\vc{p})) \union 
  \cl{C}_i(P_2,\hatt{A}(\vc{p})) \\[1ex]

  (P_1\mid P_2) &:\cl{C}_i(P_1,\hatt{A}(\vc{p})) \union
  \cl{C}_i(P_2,\hatt{A}(\vc{p})) \\[1ex]

  \new{s}{P'}   &:\cl{C}_i(P',\hatt{A}(\vc{p})) \\[1ex]

\emit{s}{e},i=0,1&:\emptyset \\[1ex]
   \_\_,i=2          &\{\hatt{A}(\vc{p})_{\infr{\rho}}  \geq_2
  e\} \qquad s:Sig_{\rho}(t) \\[1ex]

  B(\vc{e}),i=0
&:\left\{
\begin{array}{ll}
\emptyset
&\mbox{if }A>_F B \\[1ex]
\set{\hatt{A}(\vc{p}) >_0 \hatt{B}(\vc{e},\vc{y}_B)} 
&\mbox{otherwise}
\end{array}\right.\\[1ex]

\_\_,i=1 
&: \{A^+(\vc{p})_{I_A} \geq_1 B^+(\vc{e},\vc{y}_B)_{I_B}\}\\[1ex]

\_\_,i=2 
&: \left\{
    \begin{array}{ll}
      \set{A^+(\vc{p})_{\infr{\rho}} \geq_2 B^+(\vc{e},\vc{y}_B)_{\infr{\rho}} \mid
      \rho\in \mathcal{W}(B)}
      &B \not\in \w{Reset}\\[1ex]
      \emptyset&\mbox{otherwise}
    \end{array}
    \right.\\[1ex]

  \present{s^{y}}{x}{P'}{B(\vc{r})},i=0
&:\cl{C}_0([y/x]P',\hatt{A}(\vc{p})) \\[1ex]

  \_\_,i=1
&: \cp_1([y/x]P',A^+(\vc{p}))  \cup 
\{A^+(\vc{p})_{I_A} \geq_1 B^+(\vc{\ol{r}},\vc{y}_B)_{I_B}\}\\[1ex]

  \_\_,i=2 
&: \cp_2([y/x]P',A^+(\vc{p}))\\[1ex]
&\ \cup \left\{
    \begin{array}{ll}
      \set{A^+(\vc{p})_{\infr{\rho}} \geq_2
        B^+(\ol{\vc{r}},\vc{y}_B)_{\infr{\rho}} \mid
      \rho \in \mathcal{W}(B)}
      &B \not\in \w{Reset}\\[1ex]
      \emptyset&\mbox{otherwise}
    \end{array}
    \right.
\end{array}
\]
\caption{Inequalities of index $0,1,2$}\label{inequalities}
\end{table}

\subsubsection{Inequalities for termination of the instants}
In our model, the only way a computation may fail to be reactive is that a thread
goes through a recursive call infinitely often within an instant.
To avoid this situation, we have to make sure that whenever the
identifiers $A_1,\ldots,A_n$ may call each other,
a certain well-founded measure decreases. This is the purpose
of the inequalities of index $0$. Moreover, the inequalities will
be interpreted so as to make sure that a decrement step
can only be taken polynomially many times in the size of the values.

\begin{example}\label{index0-ex}
We rely on the call graphs computed in example
\ref{call-graph-ex}.
For the example \ref{cellular-automaton-ex}, we obtain:
$\hatt{\w{Send}}(s,q,\ell,\s{cons}(s',\ell''),y) >_0 \hatt{\w{Send}}(s,q,\ell,\ell'',y)$,
for the example \ref{server-ex}, we obtain:
$\hatt{\w{Handle}}(s,\s{cons}(\s{req}(s',x),\ell'))>_0
\hatt{\w{Handle}}(s,\ell')$, and 
for the example \ref{ABC-ex}, we obtain:
$\hatt{B}(s,\s{cons}(n,\ell'))
>_0
\hatt{B}(s,\ell')$.
\end{example}

\subsubsection{Inequalities for size control at the beginning of a  cycle}
The purpose of the inequalities of index $1$ is to ensure that the
size of the parameters of a thread at the {\em beginning of a new
cycle} is bounded by a function (a polynomial) of the size of the
initial parameters of the computation. Of course, a cycle starting
with $A$ may span several instants and may go through several 
recursive calls before a new cycle is started
again. For this reason, the {\em invariant} we have to maintain 
concerns {\em all} recursive calls both {\em within} and {\em at the
  end}  of the instant.

\begin{example}\label{index12-ex}
We rely again on the computation of the call graphs in example
\ref{call-graph-ex}.
For example \ref{cellular-automaton-ex}, 
assuming $I_{\w{Cell}}= \set{1,2,3}$ and 
$I_{\w{Send}}=\set{1,2,3,4}$ we obtain:
\[
\begin{array}{c}
\hatt{Cell}(s,q,\ell,0) \geq_1 \hatt{Send}(s,q,\ell,\ell,0),  \quad
\hatt{Send}(s,q,\ell,\ell',0) \geq_1 \hatt{Cell}(s,\w{next}(q,y),\ell,0), \\
\hatt{Send}(s,q,\ell,\s{cons}(s',\ell''),0) \geq_1 \hatt{Send}(s,q,\ell,\ell'',0)~.
\end{array}
\]
For example \ref{server-ex}, assuming
$I_{\w{Server}} = I_{\w{Handle}}=\set{1}$
we obtain:
\[
\begin{array}{c}
\hatt{\w{Server}}(s,0) \geq_1 \hatt{\w{Handle}}(s,0), \qquad
\hatt{\w{Handle}}(s,0) \geq_1 \hatt{\w{Handle}}(s,0), \\
\hatt{\w{Handle}}(s,0) \geq_1 \hatt{\w{Server}}(s,0)~.
\end{array}
\]
For example \ref{ABC-ex}, assuming
$I_A=I_B= I_C=\set{1}$,
we obtain:
\[
\begin{array}{c}
\hatt{A}(s,0) \geq_1 \hatt{B}(s,0),\quad
\hatt{B}(s,0) \geq_1 \hatt{B}(s,0),\quad
\hatt{B}(s,0) \geq_1 \hatt{A}(s,0),\quad
\hatt{C}(s) \geq_1 \hatt{C}(s)~.
\end{array}
\]
\end{example}

\subsubsection{Inequalities for size control within a cycle}
Finally, the purpose of the inequalities of index $2$
is to ensure that the size of any value
emitted during a cycle in a given region 
as well as the number of these emissions within an instant is polynomial 
in the size of the parameters at the beginning of the cycle,
the inputs provided by the environment, and the size of the values
read in regions of smaller rank.

\begin{enumerate}
\item 
  Given a thread identifier $A$, we compute an over approximation
  of the set of regions associated with an output within a cycle
  starting from $A$. To this end, we use the call graph defined in 
  section \ref{read-once} and we compute all thread identifiers
  that are reachable from $A$ within a cycle. Then we inspect the
  definition of each thread identifier (different from $O$) and 
  determine the regions associated  with the emissions that
  may arise in the definition. We denote with $\mathcal{W'}(A)$ this set. 
  Moreover, let $\rho_\top$ be a region whose rank is higher than the
  rank of all the regions used in the program and  
  let $\mathcal{W}(A)$ equal $\set{\rho_\top}$ if
  $\mathcal{W'}(A)=\emptyset$, and $\mathcal{W'}(A)$
  otherwise.

\item  Let $A$ be a thread identifier of arity
  $n$ with auxiliary parameters $\vc{y}_A=y_1,\ldots,y_m$.
We can associate with each position $1,\ldots,m$ in the 
list of auxiliary parameters a unique
region $\gamma(i)$ which is the region associated with the corresponding
read instruction.
Given a region $\rho$, we denote with
  $\infr{\rho}$ the set of regions of rank smaller than $\rho$.
  In particular, if $\w{rank}(\rho)=0$ then we $\infr{\rho}=\emptyset$.
  Given a set $M$ of regions we introduce the notation
  $\hatt{A}(\vc{p})_{M}$ for $\hatt{A}(\vc{p})_I$ where
  $I=\{1,\ldots,n\}\cup\{n+i \mid \gamma(i) \in M\}$.
  Thus, this amounts to set to $0$ all auxiliary parameters
  whose region is not in $M$.   Note that this masking only
  affects the {\em auxiliary} parameters of the thread identifiers.
\end{enumerate}

\begin{example}\label{region-ex}
Consider example \ref{cellular-automaton-ex}, assuming all the signals
on which the automata interact  belong to the same region $\rho$.
In this case, $\cl{W}(\w{Cell})=\cl{W}(\w{Send}) = \set{\rho}$ and
the resulting inequalities are:
\[
\begin{array}{c}
\hatt{Cell}(s,q,\ell,0) \geq_2 \hatt{Send}(s,q,\ell,\ell,0),  \\
\hatt{Send}(s,q,\ell,\s{cons}(s',\ell''),0) \geq_2 \hatt{Send}(s,q,\ell,\ell'',0),\quad
\hatt{Send}(s,q,\ell,\s{cons}(s',\ell''),0) \geq_2 q~.
\end{array}
\]
Next consider example \ref{server-ex}, assuming the region $\rho$ of the signal
on which the \w{Server} receives the requests is below the region
$\rho'$ of the signals on which it provides an answer.
In this case, 
$\cl{W}(\w{Server})=\cl{W}(\w{Handle})=\set{\rho'}$ and 
the resulting inequalities are:
\[
\begin{array}{c}
\hatt{\w{Server}}(s,y) \geq_2 \hatt{\w{Handle}}(s,y), \quad

\hatt{\w{Handle}}(s,\s{cons}(\s{req}(s',x),\ell')) \geq_2 \hatt{\w{Handle}}(s,\ell'), \\

\hatt{\w{Handle}}(s,\s{cons}(\s{req}(s',x),\ell')) \geq_2 f(x)~.

%
\end{array}
\]
Finally, consider example \ref{ABC-ex}. 
Here we have just one signal belonging, say,
to a region $\rho$.
In this case, $\cl{W}(A)= \cl{W}(B)=\cl{W}(C)=\set{\rho}$ and
the resulting inequalities are:
\[
\begin{array}{c}
\hatt{A}(s,0) \geq_2 \hatt{B}(s,y),\quad
\hatt{B}(s,\s{cons}(n,\ell')) \geq_2 \hatt{B}(s,\ell'), \quad
\hatt{B}(s,\s{cons}(n,\ell')) \geq_2 n,\\
\hatt{C}(s) \geq_2 \hatt{C}(s), \quad
\hatt{C}(s) \geq_2 n ~.
\end{array}
\]
We anticipate that the inequality $\hatt{A}(s,0) \geq_2 \hatt{B}(s,y)$
is not going to be satisfiable since
$A$ does not depend on $y$ which is a list of arbitrary size.
\end{example}

\subsection{Assignments and quasi-interpretations}\label{inequalities-sec}
We introduce first the notion of {\em assignment} which 
interprets the inequalities in terms of certain numerical functions.
A {\em quasi-interpretation} is then an assignment that {\em satisfies}
the inequalities associated with the program.

\subsubsection{Assignments}
Let $h$ denote either a constructor $\s{c}$ or a function symbol $f$ or a
thread identifier $\hatt{A}$. An {\em assignment} associates with each symbol
$h$ of arity $n$ of the program a function $q_{h}: \mathbb{N}^n \rightarrow
\mathbb{N}$ subject to a series of conditions that we specify below.

First we have to introduce some notation.
Let $E$ denote a formula which is either an expression $e$ or 
the application of a thread identifier to expressions
$\hatt{A}(e_1,\ldots,e_n)$.
Suppose $E$ contains the  variables $x_1,\ldots,x_n$. 
Once an assignment is fixed,
we can associate with $E$
a function over the natural numbers of arity $n$ by defining
$q_{x_{i}}=x_{i}$ and 
$q_{h(e_1,\dots,e_n)} = q_{h}(q_{e_{1}},\ldots,q_{e_{n}})$.
In particular, we note that if $v$ is a value then $q_v$ 
is a numerical constant.

A {\em ground substitution} is a substitution that associates
values with variables (while respecting the types).
Given two formulae $E_1,E_2$, we write $q\models E_1> E_2$
($q\models E_1 \geq E_2$) if for all ground substitutions
$\sigma$, $q_{\sigma E_{1}} > q_{\sigma E_{2}}$ 
($q_{\sigma E_{1}} \geq q_{\sigma E_{2}}$).\footnote{Sometimes, 
a stronger definition of satisfaction is considered that requires {\em e.g.}, 
$q_{E_{1}} \geq q_{E_{2}}$ where $q_{E_{1}},q_{E_{2}}$ are regarded 
as functions over the natural numbers. 
We prefer the definition based on ground substitutions 
because it allows to exploit some information on the data size. 
For instance, we may satisfy a constraint $f(x) \geq \s{c}(x,y)$ 
if we know that all the values that may replace $y$ have bounded size. 
On the other hand, with the stronger definition such constraint 
cannot be satisfied.}

We will also compare {\em vectors} of formal expressions.
For {\em lexicographic comparison}, we write
$q \models (E_1,\ldots,E_n) >_{lex} (E'_1,\ldots,E'_n)$ if
there is an $i \leq n$ such that 
$q\models E_j \geq E'_j$ for $j=1,\ldots,i-1$ and 
$q\models E_i > E'_i$.
For {\it{multi-set comparison}}, we write
$q \models (E_1,\ldots,E_n) >_{mul} (E'_1,\ldots,E'_n)$ if
for all ground substitutions $\sigma$, 
$\mset{q_{\sigma E_1},\ldots,q_{\sigma E_n}} 
>_{mset}^{\mathbb{N}} 
\mset{q_{\sigma E'_1},\ldots,q_{\sigma E'_n}} $, where
$\mset{\ldots}$ is our notation for multi-sets and
$>_{mset}^{\mathbb{N}}$ is the well-founded multi-set order over the
finite multi-sets of natural numbers. We notice the following
simple combinatorial fact about lexicographic and multi-set orders
which is instrumental to establish {\em polynomial time} termination.

\begin{lemma}\label{lex-mset-lemma}
Suppose $a_1,\ldots,a_n,c$ are natural numbers and $a_1,\ldots,a_n < c$.
Then the length of any strictly decreasing sequence of the shape
$(a_1,\ldots,a_n) >_{lex} (b_1,\ldots,b_n) >_{\w{lex}} \cdots$ or of
the shape
$\mset{a_1,\ldots,a_n} >_{mset}^{\mathbb{N}} \mset{b_1,\ldots,b_n}
>_{mset}^{\mathbb{N}} \cdots$ is bounded by $c^n$.
\end{lemma}

\begin{definition}
An assignment should satisfy the following conditions.

\Defitem{(1)} If $s$ is a signal name or $\s{c}$ is a constructor with
arity $0$ then $q_{s}=q_{\s{c}}=0$. Otherwise, if $\s{c}$ is a
constructor with positive arity $n$ 
then $q_{\s{c}}= d_{\s{c}}+\Sigma_{i=1,\ldots,n} x_i$ for 
some natural number $d_{\s{c}}\geq 1$.  

\Defitem{(2)} For all symbols $h$ of arity $n$ it holds that:
(i) $q\models h(x_1,\ldots,x_n) \geq x_i$ for $i=1,\ldots,n$ and
(ii) $q_h$ is monotonic, {\em i.e.}, $a_j\geq b_j$ for $j=1,\ldots,n$
implies $q_h(a_1,\ldots,a_n) \geq q_h(b_1,\ldots,b_n)$.

\Defitem{(3)} Let $f$ be a function symbol of arity $n$. Then
$f(v_1,\ldots,v_n) \eval v$ implies that
$q_{f(v_1,\ldots,v_n)} \geq q_{v}$.

\end{definition}

It follows from condition (1) that there is a constant $k\geq 1$ (that
can be taken as the largest additive constant $d_{\s{c}}$) such that
for any value $v$, $|v| \leq q_v \leq k\cdot |v|$.  Also note that condition (1)
implies condition (2) for constructors.

The definition of an assignment $q$
ensures that $\sigma e \Downarrow v$ implies $q_{\sigma e} \geq q_v$.
We say that a function $U:\mathbb{N} \rightarrow \mathbb{N}$ bounds
the assignment $q$ if for all symbols $h$ and all natural numbers $n$
it holds that $q_h(n,\ldots,n) \leq U(n)$.  We say that an assignment
is polynomially bounded if it can be bound by a function $U$ which is
a polynomial.  In the following, we will restrict our attention to
polynomially bounded assignments.

\subsubsection{Quasi-interpretations}
A quasi-interpretation is a polynomially bounded assignment
which satisfies the constraints of index $0,1,2$.

\begin{definition}[quasi-interpretation]
An assignment $q$ is a quasi-interpretations if:

\Defitem{(1)} For all constraints of the shape 
 $A^+(p_1,\ldots,p_n) >_0 B^+(e_1,\ldots,e_n)$ where $A=_F B$, 
with status $st$, we have:
\[
  q \models (p_1,\ldots,p_n) >_{st} (e_1,\ldots,e_n)~.
\]

\Defitem{(2)} For all constraints of the shape
$\hatt{A}(p_1,\ldots,p_n) \geq_i \hatt{B}(e_1,\ldots,e_m)$ ($i=1,2$)
and $\hatt{A}(p_1,\ldots,p_n) \geq_2 e$  we have:
\[
q \models \hatt{A}(p_1,\ldots,p_n) \geq \hatt{B}(e_1,\ldots,e_m) 
\mbox{ and }
q \models \hatt{A}(p_1,\ldots,p_n) \geq e ~.
\]
\end{definition}

\begin{example}\label{qi-ex}
Consider example \ref{cellular-automaton-ex} and assume
that we attribute the lexicographic status to $\w{Cell}$ and $\w{Send}$.
We note that $\w{Cell}>_F \w{Send}$. The inequality of index $0$
is satisfied because the quasi-interpretation of
$\s{cons}(s',\ell'')$ is always strictly larger than the quasi-interpretation
of $\ell''$.
To satisfy the remaining inequalities of index $1,2$ it suffices
to interpret $\hatt{\w{Cell}}$ and $\hatt{\w{Send}}$ as
the maximum function, noticing that $\w{next}(q,y)$ is always
a state which is represented by a constant of size $0$.

Next, consider example \ref{server-ex} and assume lexicographic
status for the thread identifiers.
We note that $\w{Handle} >_F \w{Server}$.
Again the inequality of index $0$ is satisfied because
the quasi-interpretation of 
$\s{cons}(\s{req}(s',x),\ell')$ is always larger than the
quasi-interpretation of $\ell'$.
To satisfy the inequalities of index $1,2,3$ it suffices
to suppose that the quasi-interpretation of $\hatt{Handle}$
and $\hatt{\w{Server}}$ is a function $g:\Nat^2\arrow \Nat$
such that $g(0,x)$ is pointwise larger than the quasi-interpretation
of the function $f$.
Finally, consider example \ref{ABC-ex}. We note that $A>_F B$.
We can satisfy the inequalities of index $0,1$ but as anticipated
there is no way the inequality $\hatt{A}(s,0)\geq_2 \hatt{B}(s,y)$
can be satisfied since $y$ ranges over lists of arbitrary size.
\end{example}

We can now state our main result whose proof will be discussed
in the following section \ref{proof}.

\begin{theorem}\label{main-thm}
A program that admits a polynomial quasi-interpretation is
feasibly reactive.
\end{theorem}

\section{Proofs outline} \label{proof}
We are given a finite system of recursive equations.  The initial configuration
of a program relatively to such a system has the shape: $R=\nu \vc{s}\
(A_1(\vc{v}_{1}) \mid \cdots \mid A_n(\vc{v}_{n}))$.  Since we have
assumed that the system is {\em finite control} during the computation
we will have at most $n$ main parallel threads plus a variable number
of auxiliary threads that may just branch and emit signals and that
disappear at the end of each instant. Of course one of our goals is to
show that this variable number of threads can be polynomially
bounded. 

\begin{lemma}\label{termination}
Let $R$ be a program admitting a polynomial quasi-interpretation.
There is a polynomial $Q(x)$ such that if $c$ bounds the size of $R$,
the size of the inputs, and the sizes of the parameters of all calls
within a given instant then the program in that instant will suspend in
time less than $Q(c)$.
\end{lemma}

The computation performed by the program is simply the interleaving of
the computations performed by the $n$ main threads.  It is clear that
the computation a thread may perform within an instant before running
a recursive call is polynomially bounded in $c$. Thus it is enough to
show that each thread may perform at most polynomially many recursive
calls before suspending and to this end we rely on the inequalities of
index $0$ and the lemma \ref{lex-mset-lemma}.  Note that the size and
the number of the values emitted during the instant is polynomial in
$c$ and that therefore their concatenation in a list has size
polynomial in $c$ too. 
We anticipate that the proof we have sketched of the lemma \ref{termination} 
actually shows that each thread whose parameters and
inputs are bound by $c$ will suspend in time polynomial in $c$.

\begin{lemma}\label{nsi}
Let $R$ be a program admitting a polynomial quasi-interpretation.
There is a polynomial $Q(x)$ such that
if $c$ bounds the size of $R$ and $A \in Reset$ then,
in all computations of $R$, the sizes of the parameters
in every call to $A$ are bounded by $Q(c)$.
\end{lemma}

The inequalities of index $1$ guarantee that a computation that
starts with $B(\vc{v})$ will have the property that 
$B(\vc{v})$ will `dominate' (up to quasi-interpretation and modulo the
parameter annotations) all the following calls $A(\vc{u})$ including
those that correspond to a reset point and in this case all
parameters of the call are taken into account by the definition of $I_A$.

\begin{lemma}\label{size-bound}
Let $R$ be a program admitting a polynomial quasi-interpretation.
There exists a polynomial $Q(x)$ such that in every computation
  $$R \equiv
  R_{i_0+1} \trarrow R_{i_1} \mapstoin{Env_1} R_{i_1+1}
  \trarrow R_{i_2} \mapstoin{Env_2} R_{i_2+1} \ldots
  $$
if $c$ bounds the size of $R$ and of the inputs
$Env_1,\ldots,Env_k$ for $k \geq 0$ then
the size of every value computed within
the instant $k$ is bounded by $Q(c)$.
\end{lemma}

First of all we show by induction on the rank of a region
that the size of every value computed in that region is polynomial in $c$.

If the region $\rho$ has rank $0$ the inequalities of index $2$ (in
the case where all auxiliary parameters are set to $0$) guarantee that
(i) the size of an emitted value and (ii) the size of a parameter in a
recursive call to a thread identifier that may emit on the region
$\rho$ is polynomial in the parameters at the beginning of a
cycle.  Now, by lemma \ref{nsi}, the size of the parameters 
at the beginning of a cycle is polynomial in $c$.
Thus from (the proof of) lemma \ref{termination}, we can derive
that the number of values emitted is polynomial in $c$.
We can then conclude that all the values emitted or computed at the end
of the instant by list concatenation have a size that is polynomial in
$c$.

Next suppose the region $\rho$ has rank greater than $0$.  This time
the inequalities of index $2$ (in the case where we restrict the
auxiliary parameters to those that depend on regions of rank strictly smaller
than $\rho$) guarantee that (i) the size of an emitted value and (ii)
the size of a parameter in a recursive call to a thread identifier
that may emit on the region $\rho$, is polynomial in the size of the
parameters at the beginning of a cycle {\em and} the values read from
regions strictly smaller than $\rho$.  Using the fact that the
composition of polynomials is again a polynomial we can appeal
again to  lemmas \ref{termination} and  \ref{nsi}
to conclude that all values emitted or computed at the end
of the instant by list concatenation in the region $\rho$ 
have a size that is polynomial in $c$.

There is one situation that remains to be considered.  The computation
may reach a thread identifier that that does not emit any value within
its current cycle.  By lemma \ref{termination}, it is enough to make
sure that the size of its parameters is polynomial in $c$. This is
guaranteed again by the inequalities of index $2$ since by convention
a region with the largest rank is in $\cl{W}(B)$.

Thus we have shown that the size of the values is polynomial
in the size of the initial configuration and the size of the
largest input.  By applying again lemma \ref{termination} we
can conclude that the program is feasibly reactive.

\section{Conclusion}
We have introduced the property of feasible reactivity in the context
of a synchronous $\pi$-calculus and we have provided static conditions
that enforce it.  The {\em read-once condition} builds on the cyclic
behaviour of typical synchronous applications and allows to regard
each thread as a function of its parameters and of the finitely many
inputs it receives within a cycle.  Reactivity is obtained as usual
through a {\em well-founded measure}.  In our case, this measure is
tuned so as to ensure termination in time polynomial in the size of
the values. Feasible reactivity requires that we control both the
number and the size of the threads. This is achieved in particular by
requiring that each thread at the beginning of a cyle is {\em non-size
increasing}.  To escape certain circular situations, a final condition
requires a {\em stratification of the signals in regions} so that,
intuitively, a value emitted on a certain region can be polynomially
bounded in the size of the values read in lower regions.

Various directions for further research can be mentioned.
First, it is clear that an automatisation of our approach relies
on the possibility of {\em synthesizing} quasi-interpretations.
Preliminaries experiences suggest that quasi-interpretations
are not too hard to find in practice (see, {\em e.g.}, \cite{Amadio04}),
but it remains to be seen whether this approach scales up to large programs.
Second, one might wonder whether the read-once condition can
be dropped. Currently, it plays an essential role
in the proofs and its eventual removal seems to require new ideas on
the abstraction of threads' execution. 
Third, our analysis is tailored towards the synchronous model
and a signal based interaction mechanism. It remains to be
seen whether similar analyses could be performed on different models
of concurrent threads.  For instance a model based on {\em shared references}
and possibly {\em asynchronous} execution.

\newpage

\appendix

\section{Proof of lemma \ref{lex-mset-lemma}}
Suppose $a_{n-1},\ldots,a_{0}$ are natural numbers strictly smaller
than a constant $c$. We define
\[
B_{\w{lex}}(a_{n-1},\ldots,a_{0})(c) = \Sigma_{i=0,\ldots,n-1} a_i c^i
\]
which is simply the value in base $c$ of the sequence $(a_{n-1},\ldots,a_{0})$.
We also define
\[
B_{\w{mset}}(a_{n-1},\ldots,a_{0})(c) = \Sigma_{i=0,\ldots,n-1} a_{\pi(i)} c^i
\]
where $\pi$ is a permutation over $\set{0,\ldots,n-1}$ such that
$a_{\pi(0)} \leq \cdots \leq a_{\pi(n-1)}$. The
permutation $\pi$ is not uniquely determined but the definition
of $B_{\w{mset}}$ does not depend  on its choice.

Now suppose $a_{n-1},\ldots,a_{0},b_{n-1},\ldots,b_{0}$ are natural
numbers strictly smaller than a constant $c$ and note that
$B_{\w{st}}(a_{n-1},\ldots,a_{0})(c)<c^n$ for $\w{st}\in \set{\w{lex},\w{mset}}$.
If $(a_{n-1},\ldots,a_{0}) >_{\w{lex}} (b_{n-1},\ldots,b_{0})$ then
clearly $B_{\w{lex}}(a_{n-1},\ldots,a_{0})(c) >
B_{\w{lex}}(b_{n-1},\ldots,b_{0})(c)$. Therefore, the length of a decreasing
sequence with respect to the lexicographic order is bounded by $c^n$.

On the other hand, suppose 
$M=\mset{a_{n-1},\ldots,a_{0}} >_{\w{mset}}
\mset{b_{n-1},\ldots,b_{0}}=N$.
Also assume  that $\pi,\pi'$ are permutations such that
$a_{\pi(0)} \leq \cdots \leq a_{\pi(n-1)}$ and 
$b_{\pi'(0)} \leq \cdots \leq b_{\pi'(n-1)}$.
By definition of the multi-set order, we know that there 
is a non-empty multi-subset of $M$ whose largest element is, say,
$a$ which is replaced in $N$ by another multi-set (with the same cardinality) whose
largest element is strictly smaller than $a$. For instance,
$\mset{1,2,5,5,5,7} >_{\w{mset}} \mset{4,4,4,4,5,7}$ and
$\mset{1,2,5,5}$ is replaced by $\mset{4,4,4,4}$.
Then for some $k\in \set{0,\ldots,n-1}$ we have:
$a_{\pi(n-1)}=b_{\pi'(n-1)},\ldots, a_{\pi(k+1)}=b_{\pi'(k+1)},
a= a_{\pi(k)}> b_{\pi'(k)}$. If follows that 
$B_{\w{mset}}(a_{n-1},\ldots,a_{0})(c) > 
B_{\w{mset}}(b_{n-1},\ldots,b_{0})(c)$ and again 
the length of a decreasing
sequence with respect to the multi-set order is bounded by $c^n$.

\section{Abstraction}\label{rew-rules}
We are given a finite system of recursive equations.  
Our goal is to analyse the possible computations of 
a program whose initial shape is
$R=\nu \vc{s}\ (A_1(\vc{v}_{1}) \mid \cdots \mid A_n(\vc{v}_{n}))$.
We will assume that initially all thread identifiers are reset points,
{\em i.e.}, $A_1,\ldots,A_n\in \w{Reset}$.
We will proceed in two steps. First, we will {\em abstract} the program 
(the system of equations, actually) as a {\em term-rewriting system}.
Second, we will show that the inequalities 
we have produced in table \ref{inequalities} guarantee
feasible reactivity for the abstracted system and 
therefore for the concrete one.

\subsection{Abstracting signal names}
The only information we will keep of a signal name is its type
$\w{Sig}_{\rho}(t)$.  Thus we know its region $\rho$ and the type of
the values it may carry.  Formally, we select a distinct canonical
{\em constant}, say $s$, for every type $\w{Sig}_{\rho}(t)$ and
replace in the program every occurrence of a signal name of the same
type with $s$.  Following this operation, 
we remove all name generation instructions $\nu s$.
As for the operation $[s_1=s_2]P_1,P_2$ that compares signal names,
we will simply disregard it and systematically explore the situations
where one of the programs $P_1$ or $P_2$ is executed. This is like 
replacing a conditional $[s_1=s_2]P_1,P_2$ with an internal choice 
$P_1\oplus P_2$.

\subsection{Abstracting pattern matching}
Consider a pattern matching instruction $\matchv{x}{p}{P_{1}}{P_{2}}$. 
As in the name comparison operation, we will systematically consider
the situations where $P_1$ or $P_2$ are executed. However, in the case
where the first branch $P_1$ is selected, we will remember that 
$x$ must match the pattern $p$.

\subsection{Abstracting the input}
In section \ref{annotation-sec}, we have associated a distinct label (a variable) $y$ with
every input. We rely on this variable to compute `abstractly' beyond an input.
Namely, in the input operations, say, $s^y(x).P,A(f(!^{y'}s'))$ we will consider
both the possibility where a signal is received on $s$ and the computation
continues within the instant with $[y/x]P$ and the possibility that the computation
suspends and resumes in the following instant with $A(f(y'))$.

\subsection{Rewriting rules}
We will rely on rewriting rules of the shape
\begin{equation}\label{rule-emission}
\hatt{A}(\vc{p}) \arrow \emit{s}{e} 
\end{equation}
to express the situation where 
the thread identifier $A$ with parameters and inputs
that match the patterns $\vc{p}$ emits within the same instant  
the value resulting from
the evaluation of the expression $e$ on the signal $s$.

We will also rely on rewriting rules of the shape:
\begin{equation}\label{rule-continuation}
\hatt{A}(\vc{p}) \leadsto T
\end{equation}
to describe the situation where 
the thread identifier $A$ with parameters and inputs
that match the patterns $\vc{p}$ evolves into a continuation $T$.
Here, the reduction symbol $\leadsto$ can be either $\arrow$ or
$\mapsto$ with the convention that we use $\arrow$ to describe a
situation where the continuation $T$ runs in the same instant and
$\mapsto$ to describe a situation where the continuation $T$ runs in
the following instant. 

Moreover, the continuation $T$ can have two shapes:
\begin{itemize}
\item Either $B\notin\w{Reset}$, $T=\hatt{B}(\vc{e},\vc{y}_B)$,   and $\vc{p}= \vc{p'},\vc{y}_B$, 
\item or  $B\in\w{Reset}$ and $T=\lambda \vc{y}_B.\hatt{B}(\vc{e},\vc{y}_B)$.
\end{itemize}
Thus the rule (\ref{rule-continuation}) is declined 
into {\em four} cases: the continuation $T$ can run in the same instant 
or not and it can be a reset point or not.

Here the notation $\vc{e},\vc{y}_B$ (or $\vc{p'},\vc{y}_B$) 
should be understood with a grain of salt. We just mean that the parameters
can be {\em partitioned} into two groups one of which corresponds to
the auxiliary variables of the thread identifier $B$;
the parameters $\vc{y}_B$ do not necessarily follow the others.
In case $\vc{y}_B$ is empty, we will take the convention that
$\lambda \vc{y}_B$ is a dummy abstraction. 
As usual in term-rewriting, it is assumed that the variables
free in the emitted expression $e$ or the continuation $T$ are
contained in the variables in the patterns $\vc{p}$ (recalling that
the abstraction of a signal name is treated as a constant).

\subsection{Generating the rewriting rules}
Given a finite system of recursive equations, the computation
of the term rewriting rules follows quite closely the generation
of the inequalities described in table \ref{inequalities}.
Namely for each equation $A(\vc{x})=P$ we compute the function
$\cl{R}(P,\hatt{A}(\vc{x},\vc{y}_{A}))$ which is defined 
on the structure of $P$ as follows:
\[
\begin{array}{ll}
  \cl{R}(P,\hatt{A}(\vc{p}))
  &= \s{case} \ P \ \s{of} \\[1ex]
  0&:\emptyset \\[1ex]

  \matchv{x}{p}{P_1}{P_2}  &:\cl{R}(P_1,\hatt{A}([p/x]\vc{p})) \union 
  \cl{R}(P_2,\hatt{A}(\vc{p})) \\[1ex]

  [s_1=s_2]P_1,P_2   &:\cl{R}(P_1,\hatt{A}(\vc{p})) \union 
  \cl{R}(P_2,\hatt{A}(\vc{p})) \\[1ex]

  (P_1\mid P_2) &:\cl{R}(P_1,\hatt{A}(\vc{p})) \union
  \cl{R}(P_2,\hatt{A}(\vc{p})) \\[1ex]

  \new{s}{P'}   &:\cl{R}(P',\hatt{A}(\vc{p})) \\[1ex]

\emit{s}{e}     &:\set{\hatt{A}(\vc{p}) \arrow \emit{s}{e}} \\[1ex]

 B(\vc{e})      &:\left\{
\begin{array}{ll}
\set{\hatt{A}(\vc{p}) \arrow \hatt{B}(\vc{e},\vc{y}_B)}
&\mbox{if }B\notin \w{Reset} \\
\set{\hatt{A}(\vc{p}) \arrow \lambda \vc{y}_B.\hatt{B}(\vc{e},\vc{y}_B)}
&\mbox{if }B\in \w{Reset}
\end{array}\right.\\[2ex]

\present{s^{y}}{x}{P'}{B(\vc{r})} 
&: \left\{
\begin{array}{ll}
\cl{R}([y/x]P',\hatt{A}(\vc{p}))  \union 
\set{\hatt{A}(\vc{p}) \mapsto \hatt{B}(\vc{\ol{r}},\vc{y}_B)} 
&\mbox{if }B\notin \w{Reset} \\

\cl{R}([y/x]P',\hatt{A}(\vc{p}))  \union 
\set{\hatt{A}(\vc{p}) \mapsto \lambda \vc{y}_{B}.\hatt{B}(\vc{\ol{r}},\vc{y}_B)}
&\mbox{if }B\in \w{Reset}
    \end{array}
    \right.

\end{array}
\]
Here the abstracted variables $\lambda \vc{y}_{B}$ are supposed to be
fresh.
Also note that by the shape of the rules we can never rewrite
an emission $\emit{s}{e}$ or an abstraction  such as 
$\lambda \vc{y}_{B}.\hatt{B}(\vc{e},\vc{y}_B)$ since these
terms never match the left-hand side of a rule.

\begin{example}
We compute the term rewriting rules associated with our running examples.
For example \ref{cellular-automaton-ex}, we derive:
\[
\begin{array}{c}

\hatt{\w{Cell}}(s,q,\ell,y) \arrow \hatt{\w{Send}}(s,q,\ell,\ell,y)  \\

\hatt{\w{Send}}(s,q,\ell,\s{cons}(s',\ell''),y) \arrow
\hatt{\w{Send}}(s,q,\ell,\ell'',y) \\

\hatt{\w{Send}}(s,q,\ell,\s{cons}(s',\ell''),y) \arrow
\emit{s'}{q} \\ 

\hatt{\w{Send}}(s,q,\ell,\ell',y) \mapsto
\lambda y'.\hatt{\w{Cell}}(s,\w{next}(q,y),\ell,y')

\end{array}
\]
For example \ref{server-ex}, we derive:
\[
\begin{array}{cc}

\hatt{\w{Server}}(s,y) \mapsto \hatt{\w{Handle}}(s,y) 

&\hatt{\w{Handle}}(s,\s{cons}(\s{req}(s',x),\ell')) 
\arrow \hatt{\w{Handle}}(s,\ell') \\

\hatt{\w{Handle}}(s,\s{cons}(\s{req}(s',x),\ell')) 
\arrow \emit{s'}{f(x)} 

&\hatt{\w{Handle}}(s,\ell) \mapsto \lambda y.\hatt{\w{Server}}(s,y)

\end{array}
\]
Finally, for example \ref{ABC-ex}, we derive:
\[
\begin{array}{cc}

\hatt{A}(s,y) \mapsto \hatt{B}(s,y) 

&\hatt{B}(s,\s{cons}(n,\ell')) 
\arrow \emit{s}{n} \\

\hatt{B}(s,\s{cons}(n,\ell')) 
\arrow \hatt{B}(s,\ell') 

&\hatt{B}(s,\ell) \mapsto \lambda y.\hatt{A}(s,y) \\

\hatt{C}(s)      \arrow \emit{s}{n} 

&\hatt{C}(s)      \mapsto \lambda ().\hatt{C}(s)

\end{array}
\]
\end{example}

\begin{remark}
The reader might have noticed that the rewriting rules 
and the inequalities we have produced do not keep track of 
events that can happen in {\em parallel} like
``emitting two signals and calling another thread''.
This information can be neglected because we have assumed
we are handling {\em finite control} programs. In such programs
a call to an identifier $A$ may generate at most one
call to another thread identifier (either in the current
instant or in the following one) plus a number of
emissions that is bounded by a constant that depends on the
size of the program only.
Alternatively, we could have considered rewriting rules  such as:
\[
\hatt{A}(\vc{p}) \arrow \emit{s_{1}}{e_{1}} \ \| \
\emit{s_{2}}{e_{2}} \ \| \ \hatt{B}(\vc{e},\vc{y}_B)
\]
where the right hand side carries a composition operator $\|$ to express
the parallelism of the events.
We note that this approach may produce exponentially more rules than the previous
one because one needs to distribute the parallel composition through
the non-determinism. 
\end{remark}

\section{Analysis}
We proceed to an analysis of the abstracted system, {\em i.e.}, of 
the term rewriting system.
Table \ref{rules-inequalities} summarizes the inequalities 
that are associated with each kind of  term
rewriting rule.

\begin{table}
\[
\begin{array}{l|l}

\mbox{Rewriting Rules} & \mbox{Associated Inequalities} \\[1ex]\hline

(R1)\quad \hatt{A}(\vc{p})\arrow \emit{s}{e}, s:\w{Sig}_{\rho}(t) 

&\hatt{A}(\vc{p})_{\infr{\rho}} \geq_2 e \\[1ex]

(R2)\quad \hatt{A}(\vc{p},\vc{y}_B)\arrow \hatt{B}(\vc{e},\vc{y}_B)
&\left\{
\begin{array}{l}
\hatt{A}(\vc{p},\vc{y}_B)>_0 \hatt{B}(\vc{e},\vc{y}_B) \mbox{ if }A =_F B\\
\hatt{A}(\vc{p},\vc{0})_{I_{A}}\geq_1 \hatt{B}(\vc{e},\vc{0})_{I_{B}} \\
\hatt{A}(\vc{p},\vc{y}_B)_{\infr{\rho}} 
\geq_2 \hatt{B}(\vc{e},\vc{y}_B)_{\infr{\rho}} \mbox{ if }\rho\in \cl{W}(B) 
\end{array} \right.\\[3ex]

(R3)\quad \hatt{A}(\vc{p})\arrow \lambda \vc{y}_B.\hatt{B}(\vc{e},\vc{y}_B)
&\hatt{A}(\vc{p})_{I_{A}}\geq_1 \hatt{B}(\vc{e},\vc{0}) \\[1ex] 

(R4)\quad \hatt{A}(\vc{p},\vc{y}_B)\mapsto \hatt{B}(\vc{\ol{r}},\vc{y}_B)
&\left\{
\begin{array}{l}
\hatt{A}(\vc{p},\vc{0})_{I_{A}}\geq_1 \hatt{B}(\vc{\ol{r}},\vc{0})_{I_{B}} \\
\hatt{A}(\vc{p},\vc{y}_B)_{\infr{\rho}} 
\geq_2 \hatt{B}(\vc{\ol{r}},\vc{y}_B)_{\infr{\rho}} \mbox{ if }\rho\in \cl{W}(B) 
\end{array} \right.\\[2ex] 

(R5)\quad \hatt{A}(\vc{p})\mapsto \lambda \vc{y}_B.\hatt{B}(\vc{\ol{r}},\vc{y}_B)
&\hatt{A}(\vc{p})_{I_{A}}\geq_1 \hatt{B}(\vc{\ol{r}},\vc{0}) 

\end{array}
\]
\caption{Inequalities associated with the term rewriting rules}\label{rules-inequalities}
\end{table}

A term rewriting rule describes a family of {\em ground} rewriting
rules which is obtained by replacing the variables with ground
substitutions $\sigma$ and by evaluating the ground expressions according
to the evaluation axioms. We write
\[
\hatt{A}(\vc{v}) \act{R1} \emit{s}{v}
\]
if there is a term rewriting rule $\hatt{A}(\vc{p}) \arrow
\emit{s}{e}$ and a ground substitution $\sigma$ such that
$\sigma \vc{p} = \vc{v}$ and $\sigma e \eval v$.
In a similar way, we write 
\[
\hatt{A}(\vc{v},\vc{u}) \stackrel{R2}{\arrow} 
\hatt{B}(\vc{v'},\vc{u})
\qquad(\mbox{or}\qquad 
\hatt{A}(\vc{v},\vc{u}) \stackrel{R4}{\mapsto} 
\hatt{B}(\vc{v'},\vc{u}) \ )
\]
if there is a term rewriting rule $\hatt{A}(\vc{p},\vc{y}_B) \arrow
\hatt{B}(\vc{e},\vc{y}_B)$ (or $\hatt{A}(\vc{p},\vc{y}_B) \mapsto
\hatt{B}(\vc{\ol{r}},\vc{y}_B)$)
and a ground substitution $\sigma$ such that
$\sigma \vc{p} = \vc{v}$, $\sigma \vc{y}_B = \vc{u}$, and 
$\sigma \vc{e} \eval \vc{v'}$ (or $\sigma \vc{\ol{r}}\eval \vc{v'}$).
Finally, we write
\[
\hatt{A}(\vc{v}) \stackrel{R3}{\arrow} \lambda
\vc{y}_B.\hatt{B}(\vc{v'},\vc{y}_B)
\qquad(\mbox{or}\qquad
\hatt{A}(\vc{v}) \stackrel{R5}{\mapsto} \lambda
\vc{y}_B.\hatt{B}(\vc{v'},\vc{y}_B) \ )
\]
if there is a term rewriting rule $\hatt{A}(\vc{p}) \arrow
\lambda \vc{y}_B.\hatt{B}(\vc{e},\vc{y}_{B})$ 
(or $\hatt{A}(\vc{p}) \mapsto
\lambda \vc{y}_B.\hatt{B}(\vc{\ol{r}},\vc{y}_{B})$)
and a ground substitution $\sigma$ such that
$\sigma \vc{p} = \vc{v}$,  and  $\sigma \vc{e} \eval \vc{v'}$ 
(or $\sigma \vc{\ol{r}}\eval \vc{v'}$).

Consider a ground rewriting rule representing a computation
step.  As we have seen this rule is an instance of a term rewriting
rule.  In turn, we have associated a set of inequalities with every
term rewriting rule.  Let us now assume we have an assignment $q$ that
satisfies all generated inequalities. Table \ref{ground-qint} spells 
out what this means in terms of the ground rewriting rule.  
To this end, we need some notation
to distinguish the parameters $\vc{e}$ of a thread identifier $\hatt{A}$
(remember that a list of variables $\vc{y}_B$ or a list of patterns $\vc{p}$ 
is also a list of expressions and that 
$\vc{\ol{r}}$ is a list of expressions too since, by definition, the
dereferenced signals are replaced by variables).
We distinguish between proper parameters and auxiliary
parameters (those corresponding to an input). 
Among the former, we distinguish those in the set $I_{A}$ 
($\vc{e}_{I_{A}}$) and the others ($\vc{e}_{\ol{I_{A}}}$). 
Among the latter, for a given region $\rho$,
we distinguish those whose rank is smaller than $\rho$ 
($\vc{e}_{\infr{\rho}}$) and the others ($\vc{e}_{\ol{\infr{\rho}}}$).
To summarise, given a list of parameters $\vc{\vc{e}}$ and a region
$\rho$, 
we can always
partition it into four parts:
$\vc{e} = \vc{\vc{e}}_{I_{A}},\vc{e}_{\ol{I_{A}}}, \vc{e}_{\infr{\rho}},
\vc{e}_{\ol{\infr{\rho}}}$.

We also notice that $\hatt{A}(\vc{e},\vc{y}_{A})_{I_{A}}=\hatt{A}(\vc{e},\vc{0})_{I_{A}}$
since by definition all auxiliary parameters are set to $\s{0}$. 
Moreover, if $A$ is a reset point then $\hatt{A}(\vc{e},\vc{0})_{I_{A}} = 
\hatt{A}(\vc{e},\vc{0})$ since for the reset points, $I_{A}$ coincides with the 
proper parameters.
Finally, we recall that the restriction $\infr{\rho}$ acts only on the auxiliary 
parameters.

\begin{table}
\[
\begin{array}{lc}

(R1)&
\infer{\hatt{A}(\vc{p}) \arrow \emit{s}{e}, 
\quad s:\w{Sig}_{\rho}(t), 
\quad \vc{p}=\vc{p}_1,\vc{p}_2, 
\quad \vc{p}_2=\vc{p}_{\ol{\infr{\rho}}}, \\
\sigma (\vc{p}_{1},\vc{p}_2) = \vc{v}_{1},\vc{v}_2, 
\quad \sigma e \eval v}
{q \models \hatt{A}(\vc{v}_{1},\vc{0}) \geq v}  \\ \\

(R2)&
\infer{
\hatt{A}(\vc{p},\vc{y}_B) \arrow \hatt{B}(\vc{e},\vc{y}_B), \quad 
\rho \in \cl{W}(B),\quad 
\vc{p},\vc{y}_B=\vc{p}_1,\ldots,\vc{p}_4,\vc{y}_5,\vc{y}_6, \\
\vc{p}_{1} = \vc{p}_{I_{A}},             
\quad \vc{p}_{2} = \vc{p}_{\ol{I_{A}}}, 
\quad \vc{p}_{3} = \vc{p}_{\infr{\rho}},  
\quad \vc{p}_{4} = \vc{p}_{\ol{\infr{\rho}}},  
\quad \vc{y}_{5} = (\vc{y}_{B})_{\infr{\rho}},
\quad \vc{y}_{6} = (\vc{y}_{B})_{\ol{\infr{\rho}}}, \\
\vc{e}= \vc{e}_1,\vc{e}_2, 
\quad \vc{e}_1 = \vc{e}_{I_{B}},
\quad \vc{e}_2 = \vc{e}_{\ol{I_{B}}}\\
\sigma(\vc{p}_{1},\ldots,\vc{p}_4,\vc{y}_{5},\vc{y}_{6}) = 
\vc{v}_{1},\ldots,\vc{v}_4,\vc{u}_{5},\vc{u}_{6} =\vc{v},\vc{u},
\quad \sigma (\vc{e}_{1},\vc{e}_{2}) \eval (\vc{v'}_{1},\vc{v'}_{2})=\vc{v'}}
{
q \models (\vc{v},\vc{u}) >_{\w{st}} (\vc{v'},\vc{u}), \mbox{ if }A=_F B,\w{status}(A)=\w{status}(B)=\w{st}, \\
q \models \hatt{A}(\vc{v}_{1},\vc{0},\vc{0},\vc{0},\vc{0},\vc{0}) \geq
\hatt{B}(\vc{v'}_{1},\vc{0},\vc{0},\vc{0}),  \\

q \models \hatt{A}(\vc{v}_{1},\vc{v}_{2},\vc{v}_{3},\vc{0},\vc{u}_{5},\vc{0}) \geq
\hatt{B}(\vc{v'}_{1},\vc{v'}_{2},\vc{u}_{5},\vc{0}) 
} \\ \\

(R3) &
\infer{\begin{array}{c}
\hatt{A}(\vc{p}) \arrow \lambda \vc{y}_{B}.\hatt{B}(\vc{e},\vc{y}_B), 
\quad \vc{p}=\vc{p}_1,\vc{p}_2,
\quad \vc{p}_{1} = \vc{p}_{I_{A}},\\
\sigma (\vc{p}_{1},\vc{p}_{2}) = \vc{v}_{1},\vc{v}_{2},
\quad \sigma \vc{e}  \eval \vc{v'}
\end{array}}
{
q\models \hatt{A}(\vc{v}_{1},\vc{0}) \geq
\hatt{B}(\vc{v'},\vc{0})  
} \\ \\

(R4)&
\infer{\hatt{A}(\vc{p},\vc{y}_B) \mapsto \hatt{B}(\vc{\ol{r}},\vc{y}_B),
\quad \rho \in \cl{W}(B), 
\quad \vc{p},\vc{y}_B=\vc{p}_1,\ldots,\vc{p}_4,\vc{y}_5,\vc{y}_6, \\
\vc{p}_{1} = \vc{p}_{I_{A}},             
\quad \vc{p}_{2} = \vc{p}_{\ol{I_{A}}}, 
\quad \vc{p}_{3} = \vc{p}_{\infr{\rho}},  
\quad \vc{p}_{4} = \vc{p}_{\ol{\infr{\rho}}},  
\quad \vc{y}_{5} = (\vc{y}_{B})_{\infr{\rho}},
\quad \vc{y}_{6} = (\vc{y}_{B})_{\ol{\infr{\rho}}}, \\
\vc{\ol{r}}= \vc{\ol{r}}_1,\vc{\ol{r}}_2, 
\quad \vc{\ol{r}}_1 = \vc{\ol{r}}_{I_{B}},
\quad \vc{\ol{r}}_2 = \vc{\ol{r}}_{\ol{I_{B}}}\\
\sigma(\vc{p}_{1},\ldots,\vc{p}_4,\vc{y}_{5},\vc{y}_{6}) = 
\vc{v}_{1},\ldots,\vc{v}_4,\vc{u}_{5},\vc{u}_{6},
\quad 
\sigma (\vc{\ol{r}}_{1},\vc{\ol{r}}_{2}) \eval \vc{v'}_{1},\vc{v'}_{2}}
{
q\models \hatt{A}(\vc{v}_{1},\vc{0},\vc{0},\vc{0},\vc{0},\vc{0}) \geq
\hatt{B}(\vc{v'}_{1},\vc{0},\vc{0},\vc{0}),  \\
q\models \hatt{A}(\vc{v}_{1},\vc{v}_{2},\vc{v}_{3},\vc{0},\vc{u}_{5},\vc{0}) \geq
\hatt{B}(\vc{v'}_{1},\vc{v'}_{2},\vc{u}_{5},\vc{0}) 
}\\ \\ 

(R5)&
\infer{\hatt{A}(\vc{p}) \mapsto \lambda
 \vc{y}_{B}.\hatt{B}(\vc{\ol{r}},\vc{y}_B),  
\quad \vc{p}=\vc{p}_1,\vc{p}_2,
\quad \vc{p}_{1} = \vc{p}_{I_{A}},\\          
\sigma (\vc{p}_{1},\vc{p}_{2}) = \vc{v}_{1},\vc{v}_{2},
\quad \sigma \vc{\ol{r}}  \eval \vc{v'} }
{q \models \hatt{A}(\vc{v}_{1},\vc{0}) \geq
\hatt{B}(\vc{v'},\vc{0})}   \\ \\ 

\end{array}
\]
\caption{What the quasi-interpretation guarantees of a ground
  rewriting step}\label{ground-qint}
\end{table}

\subsection{Proof of lemma \ref{termination}}
We analyse ground reductions of the shape:
\[
\hatt{A}_{1}(\vc{v}_{1}) \act{R2} \cdots \act{R2} \hatt{A}_{k}(\vc{v}_{k})
\]
These reductions correspond to a sequence of recursive calls that happen within
the same instant (and the same cycle). 
Suppose the maximum arity of a thread identifier 
$\hatt{A}$ in a given program is $n$. 
Moreover, suppose $c$ is a bound on the size of the values $\vc{v}_{j}$ for
$j=1,\ldots,k$. Then the length $k$ of the reduction sequence is
$O(c^n)$. To see this, notice that $A_{i} \geq_F A_{i+1}$ for
$i=1,\ldots k-1$. The inequality $\geq_F$ can be  strict at most 
a constant number of times that depends on the program. 
Thus, it suffices to prove the assertion
when $A_{i} =_F A_{i+1}$, for $i=1,\ldots,k-1$ knowing
that the thread identifiers have the same status $\w{st}$ and the same
arity $n$.
By the existence of a quasi interpretation $q$ (case $(R2)$ in table \ref{ground-qint}), we have:
\[
q \models \vc{v}_1 >_{\w{st}} \vc{v}_2 >_{\w{st}} \cdots >_{\w{st}} \vc{v}_{k}
\]
By the properties of assignments, we know that the interpretation of
a value is proportional to its size. 
Thus we can conclude by applying lemma \ref{lex-mset-lemma}.

\subsection{Proof of lemma \ref{nsi}}
We analyse ground reductions of the shape:
\[
\hatt{A}_{1}(\vc{v}_{1}) \leadsto \cdots \leadsto \hatt{A}_{n}(\vc{v}_{n}) 
\leadsto \lambda \vc{y}_{A_{n+1}}.\hatt{A_{n+1}}(\vc{v},\vc{y}_{A_{n+1}})
\]
where $A_1\in \w{Reset}$, $\leadsto \in \set{\arrow,\mapsto}$, 
and the last reduction is optional.
These reductions correspond to a sequence of recursive calls
that start with a reset point and continue within a cycle
(but may span several instants). Optionally, these reductions
may reach another reset point.
Let us denote with $(\vc{v}_{j})_{I_{{A}_{j}}}$ the parameters whose
indexes correspond to $I_{{A}_{j}}$. Recall that if $B$ is a reset
point then $I_B$ coincides with the proper parameters.
By the cases $(R2)$ and $(R4)$ in table \ref{ground-qint} we have:
\[
q\models
\hatt{A}_{1}((\vc{v}_{1})_{I_{{A}_{1}}},\vc{0}) 
\geq \cdots \geq 
\hatt{A}_{n}((\vc{v}_{n})_{I_{{A}_{n}}},\vc{0})~.
\]
Moreover, at the last optional step, by inspection of the cases $(R2)$ and $(R4)$ 
in table \ref{ground-qint}, we deduce:
\[
q \models \hatt{A}_{n}((\vc{v}_{n})_{I_{{A}_{n}}},\vc{0}) \geq \hatt{A_{n+1}}(\vc{v},\vc{0})~.
\]
In other terms, we know that if starting from a call $A(\vc{v})$ 
we arrive at a call $B(\vc{u})$ then
$q\models \hatt{A}(\vc{v},\vc{0}) \geq \hatt{B}((\vc{u})_{I_{B}},\vc{0})$.
In particular, we see that, up to the quasi-interpretation, the 
initial configuration $A(\vc{v})$ dominates all the following configurations
at the beginning of a cycle.

\subsection{Proof of lemma \ref{size-bound}}
We analyse ground reductions of the shape:
\[
\hatt{A}_{1}(\vc{v}_{1}) \leadsto \cdots \leadsto \hatt{A}_{n}(\vc{v}_{n})
\act{R1}  \emit{s}{v} 
\]
where $A_1\in \w{Reset}$, $\leadsto \in \set{\arrow,\mapsto}$, 
$\rho \in \cl{W}(A_{n})$, and the last reduction $R1$ is optional 
with $s$ also belonging to region $\rho$.

These reductions correspond to a sequence of recursive calls
that start with a reset point and continue within a cycle
(but may span several instants). Optionally, these reductions
may reach a point where a value is emitted.

Let $c$ be a bound on the size of the parameters
at the beginning of the computation and the size of the
values emitted by the environment at the beginning of
an instant.
Each vector $\vc{v}_{j}$ can be decomposed in 
$\vc{v'}_{j},\vc{v''}_{j}$ where $\vc{v''}_{j}$
correspond to the auxiliary parameters on regions
whose rank is not smaller than $\rho$'s.
Applying cases $(R_1)$, $(R2)$ and $(R4)$ in table \ref{ground-qint}, we deduce:
\[
q\models 
\hatt{A}_{1}(\vc{v'}_{1},\vc{0}) \geq \cdots \geq \hatt{A}_{n}(\vc{v'}_{n},\vc{0}) \geq  v
\]
Therefore we establish:

\paragraph{Property A}
The size of the emitted value $v$ is polynomial in $c$
and the size of the values read in regions whose 
rank is smaller than $\rho$'s. \\

How many times can a value be emitted on a region $\rho$ within
an instant? Between two calls, a thread can only emit
a number of messages which is bounded by a constant.
Therefore, as in lemma \ref{termination}, 
it is enough to focus on the length of computations that
happen within an instant.
We focus on ground reductions of the shape:
\[
\hatt{A}_{1}(\vc{v}_{1}) \leadsto \cdots \leadsto \hatt{A}_{k}(\vc{v}_{k}) \act{R2} \cdots \act{R2} \hatt{A}_{n}(\vc{v}_{n})
\]
where $A_1\in \w{Reset}$, $\leadsto \in \set{\arrow,\mapsto}$,
$A_k =_F \cdots =_F A_n$, 
$\w{st}=\w{status}(A_k) = \cdots = \w{status}(A_n)$,
and $\rho \in \cl{W}(A_{j})$, for $j=k,\ldots,n$.

These reductions are a particular case of those considered above,
where we suppose that after an initial sequence of recursive calls
the computation reaches a series of calls among thread identifiers
that can mutually call each other.
Let $\vc{u}$ be the arguments that correspond to the auxiliary
parameters of $\hatt{A}_j$ for $j=k,\ldots,n$.
Each vector $\vc{v}_{j}$ can be decomposed in 
$\vc{v'}_{j},\vc{u}$ for $j=1,\ldots,n$.

Applying cases $(R2)$ and $(R4)$ in table \ref{ground-qint}, we deduce:
\[
q\models 
\hatt{A}_{1}(\vc{v}_{1})_{\infr{\rho}} \geq \cdots \geq 
\hatt{A}_{k}(\vc{v'}_{k},(\vc{u})_{\infr{\rho}},\vc{0})) \geq  
\hatt{A}_{n}(\vc{v'}_{n},(\vc{u})_{\infr{\rho}},\vc{0}))
\]

\paragraph{Property B}
The parameters $\vc{v'}_{j}$ for $j=k,\ldots,n$ 
are polynomial in $c$ and the size of values read in regions
whose rank is smaller than $\rho$'s. \\

Remember that by construction there is always a region 
$\rho$ in $\cl{W}(A_{k})$. Therefore, property B guarantees
that the size of the proper parameters of a call to a thread
identifier is under control.

Now, applying case $(R2)$ in table  \ref{ground-qint}, we deduce:
\[
q\models
(\vc{v'}_{k},\vc{u}) >_{\w{st}} \cdots >_{\w{st}}
(\vc{v'}_{n},\vc{u})
\]
Because the values $\vc{u}$ are constant, we are forced
to decrease the parameters $\vc{v'}_{j}$ with respect to the
status $\w{st}$. By Property B, these parameters
are  polynomial in $c$ and the size of the values read in regions
whose rank is smaller than $\rho$'s. By lemma \ref{termination},
we know that the length of the sequence is polynomial in the size
of the largest parameter. Thus we compose the polynomials
to obtain the following.

\paragraph{Property C}
The number of times a value can be emitted within an instant
in a region $\rho$ is polynomial in $c$ and the size of 
values read in regions whose rank is smaller than $\rho$'s. \\

It remains to  analyse how in our model the size of the values {\em read} 
from a region depends on the size of the values {\em emitted} 
in that region. We have the following property.

\paragraph{Property D}
The values read from a region $\rho$ are the concatenation
of some of the values emitted in the region $\rho$ within
the same instant. \\

We can now proceed by induction on the rank of region $\rho$ to show
that the size of the concatenation of some of the values emitted in
the region $\rho$ is polynomial in $c$.  At rank $0$, we use directly
properties A and C noticing that the concatenation of polynomially
many values whose size is polynomial in $c$ produces a value which is
again polynomial in $c$.  At rank $n+1$, we use again properties A and
C and the inductive hypothesis.  Obviously the {\em degree} of the
polynomial will depend on the highest rank of a region which depends
on the program only.

\subsection{Proof of theorem \ref{main-thm}}
We can now conclude our proof.  Since the size of the computed values
is polynomial in $c$, we can apply lemma \ref{termination} and
derive that each instant terminates in time polynomial in $c$.
Thus the existence of a quasi-interpretation entails feasible reactivity.


{\footnotesize
\begin{thebibliography}{99}
  
  
\bibitem{Amadio05} R.~Amadio.
  \newblock The SL synchronous language, revisited.
  \newblock {\em Journal of Logic and Algebraic Programming}, 70:121-150, 2007.


\bibitem{Amadio06} R.~Amadio.
  \newblock A synchronous $\pi$-calculus.
  \newblock Technical Report, Universit\'e Paris 7, Laboratoire PPS, 
  June 2006. 
  \newblock {\tt http://hal.ccsd.cnrs.fr/PPS/}.
To appear in {\em Information and Computation}.

  
\bibitem{ABBC05} R.~Amadio, G.~Boudol, F.~Boussinot and I.~Castellani.
  \newblock Reactive programming, revisited.
\newblock In Proc. Workshop on {\em Algebraic Process Calculi: the
  first $25$ years and beyond}, 
{\em Electronic Notes in Theoretical Computer Science}, 162:49-60, 2006.

  
\bibitem{Amadio04} R.~Amadio.  \newblock Synthesis of max-plus
  quasi-interpretations.  \newblock In {\em Fundamenta Informaticae},
  65(1-2):29--60, 2005. 
  
\bibitem {AD04} R.~Amadio, S.~Dal-Zilio.
  \newblock Resource control for synchronous cooperative threads.
\newblock In Theoret. Comp. Sci, 358:229-254, 2006.
  
\bibitem {AD05} R.~Amadio, F.~Dabrowski.
  \newblock Feasible reactivity for synchronous cooperative threads.
  \newblock In {\em Proc. EXPRESS}, ENTCS, 154(3), 2006, 

\bibitem{BC92} S.~Bellantoni and S.~Cook.  \newblock A new
  recursion-theoretic characterization of the poly-time functions.
  \newblock {\em Computational Complexity}, 2:97--110, 1992.

  
\bibitem{BG92} G.~Berry and G.~Gonthier, \newblock The Esterel
  synchronous programming language.  \newblock {\em Science of
    computer programming}, 19(2):87--152, 1992.
  
\bibitem{BMM01} G.~Bonfante, J.-Y. Marion, and J.-Y. Moyen.  \newblock
   On termination methods with space bound certifications.  \newblock
   In {\em Proc. Perspectives of System Informatics}, Springer LNCS
   2244, 2001.
  
  
\bibitem{BD95} F.~Boussinot and R.~De Simone, \newblock The SL
  Synchronous Language.  \newblock {\em IEEE Trans. on Software
    Engineering}, 22(4):256--266, 1996.



  
\bibitem{Cobham65} A.~Cobham.  \newblock The intrinsic computational
  difficulty of functions.  \newblock In {\em Proc.  Logic,
    Methodology, and Philosophy of Science II}, North Holland, 1965.

\bibitem{Hofmann02} M.~Hofmann.  \newblock The strength of non
  size-increasing computation. \newblock In {\em Proc. ACM-POPL}, 2002.


\bibitem{Jones97} N.~Jones.  \newblock {\em Computability and
    complexity, from a programming perspective}.  \newblock MIT-Press,
  1997.



\bibitem{Leivant94} D.~Leivant.  \newblock Predicative recurrence and
  computational complexity i: word recurrence and poly-time.
  \newblock {\em Feasible mathematics II, Clote and Remmel (eds.)},
  Birkh\"auser:320--343, 1994.

\bibitem{MandelPouzetPPDP05}
  L.~Mandel and M.~Pouzet.
  \newblock Reactive{ML}, a reactive extension to {ML}.
  \newblock In {\em Proc. ACM Principles and Practice of
    Declarative Programming}, 2005.
  

\bibitem{Milner89} R.~Milner. \newblock Communication and Concurrency.
  \newblock Prentice-Hall, 1989.
  
\bibitem{mimosarp} Reactive Programming, INRIA, Mimosa Project.
  \url{http://www-sop.inria.fr/mimosa/rp}.

\bibitem{MPW92} R.~Milner, J.~Parrow, and D.~Walker. A calculus of mobile processes, parts 1-2. {\em Information and Computation}, 100(1):1--77, 1992.


\bibitem{SchemeFT}
  M.~Serrano, F.~Boussinot, and B.~Serpette.
  \newblock Scheme fair threads.
  \newblock In {\em Proc. ACM Principles and practice of 
    declarative programming}, 2004. 
  
\end{thebibliography}}
\end{document}